\documentclass[runningheads]{llncs}

\usepackage[T1]{fontenc}
\usepackage{graphicx}
\usepackage{url}

\title{Paraconsistent Semantics for Extended Fuzzy Logic Programs via Approximation Fixpoint Theory\iflong\ {[Extended Version]}\fi}
\titlerunning{Paraconsistent Semantics for Extended Fuzzy Logic Programs via AFT}

\author{
    Pascal Kettmann\inst{1\orcidlink{0009-0009-9461-7952}}\textsuperscript{\Envelope} \and
    Hannes Strass\inst{1,4\orcidlink{0000-0001-6180-6452}} \and
    Jesse Heyninck\inst{2,3\orcidlink{0000-0002-3825-4052}} \and
    Jeroen Spaans\inst{2\orcidlink{0000-0001-7027-8102}}
}

\institute{
    TU Dresden, Germany
    \email{\{pascal.kettmann,hannes.strass\}@tu-dresden.de}
    \and
    Open Universiteit, The Netherlands
    \email{\{jesse.heyninck,jeroen.spaans\}@ou.nl}
    \and
    University of Cape Town, South Africa
    \and
    ScaDS.AI Center for Scalable Data Analytics and \iflong Artificial Intelligence\else AI\fi, Dresden/Leipzig, Germany
}

\usepackage[british]{babel}
\usepackage[babel]{microtype}
\usepackage{relsize,setspace}
\usepackage{url}
\usepackage{tikz,pgf}
\usetikzlibrary{shapes,backgrounds,arrows.meta,positioning,calc}
\usepackage{amsmath,amssymb}
\usepackage[hidelinks]{hyperref}
\usepackage{cleveref}
\usepackage{multirow}
\usepackage{comment}
\usepackage{xspace}
\usepackage{orcidlink}
\usepackage{bbding}

\spnewtheorem*{proofsketch}{Proof sketch}{\itshape}{\itshape} %
\spnewtheorem{myclaim}{Claim}{\itshape}{\itshape}
\newenvironment{claimproof}
{\noindent\textit{Proof of the claim.}}
{\par}

\renewcommand{\theauxlemma}{\theproposition.\arabic{auxlemma}}
{\hfill$\triangle$\par} %

\newcounter{longcounter}

\spnewtheorem{longlemma}[longcounter]{Lemma}{\bfseries}{\itshape}
\spnewtheorem{longproposition}[longcounter]{Proposition}{\bfseries}{\itshape}
\spnewtheorem{longcorollary}[longcounter]{Corollary}{\bfseries}{\itshape}

\Crefname{longlemma}{Lemma}{Lemmata}
\newcommand{\qedhere}{\tag*{\qed}}

\usepackage{ifthen}
\usepackage{environ}
\newif\iflong
\NewEnviron{shortproof}{\iflong\else\expandafter\begin{proof}\BODY\end{proof}\fi}
\NewEnviron{shortproofsketch}{\iflong\else\expandafter\begin{proofsketch}\BODY\end{proofsketch}\fi}
\NewEnviron{longproof}{\iflong\expandafter\begin{proof}\BODY\end{proof}\fi}

\NewEnviron{longtext}{\iflong\expandafter\BODY\fi}
\NewEnviron{shorttext}{\iflong\else\expandafter\BODY\fi}

\usepackage{todonotes}

\newcommand{\N}{\mathbb{N}}

\newcommand{\R}{\mathbb{R}}
\newcommand{\Q}{\mathbb{Q}}

\newcommand{\set}[1]{\left\{#1\right\}}

\newcommand{\guard}{\ \middle\vert\ }
\newcommand{\card}[1]{\left\vert#1\right\vert}

\newcommand{\tuple}[1]{\left(#1\right)}
\newcommand{\rtuple}[1]{\langle#1\rangle}
\newcommand{\pheq}{\mathbin{\phantom{=}\,}}
\newcommand{\phiff}{\mathbin{\phantom{\iff}\,}}
\newcommand{\tand}{\;\text{and}\;}

\newcommand{\define}[1]{\emph{#1}}

\newcommand{\ordn}{\alpha}
\newcommand{\ordnalt}{\kappa}
\newcommand{\ordsucc}{\mathord{\ordn+1}}
\newcommand{\ordlim}{\gamma}

\newcommand{\eqdef}{\mathrel{:=}}
\newcommand{\iffdef}{\mathrel{:\mkern-11mu\mathord{\iff}}}

\newcommand{\ebnfeq}{\mathrel{::=}}
\newcommand{\ebnfalt}{\mathbin{\mid}}

\providecommand{\citep}[1]{\cite{#1}}
\newcommand{\citettarskitheorem}{Tarski's fixpoint theorem~\cite{tarski55fixpoint}\xspace}
\ifdefined\citeauthor\let\oldciteauthor\citeauthor\renewcommand{\citeauthor}[2]{\oldciteauthor{#2}}\else\newcommand{\citeauthor}[2]{#1}\fi
\providecommand{\citeyearpar}[1]{\cite{#1}}

\newlength{\narrowgatherskip}
\newenvironment{narrowgather}[1][3pt]{%
    \setlength{\narrowgatherskip}{#1}%
    \vskip\narrowgatherskip\noindent\hskip0pt\hfill\(}{\)\hfill\hskip0pt\vskip\narrowgatherskip}

\newcommand{\flpcon}{f}
\newcommand{\Flpcon}{\mathcal{F}}
\newcommand{\Atoms}{\Pi}
\newcommand{\atom}{p}
\newcommand{\fint}{I}
\newcommand{\fintsup}[2][i]{\fint^{(#1,#2)}}
\newcommand{\fintalt}{J}
\newcommand{\fintaltsup}[2][i]{\fintalt^{(#1,#2)}}
\newcommand{\fintset}{\mathcal{J}}
\newcommand{\lint}{L}
\newcommand{\lintsup}[1][i]{\lint^{(#1)}}
\newcommand{\uint}{U}
\newcommand{\uintsup}[1][i]{\uint^{(#1)}}
\newcommand{\lintalt}{M}
\newcommand{\uintalt}{T}

\newcommand{\nint}{K}
\newcommand{\nintalt}{L}
\newcommand{\Nint}{\mathcal{\nint}}
\newcommand{\eint}{G}
\newcommand{\eintalt}{H}
\newcommand{\Eint}{\mathcal{\eint}}
\newcommand{\dom}[1][\eint]{\mathrm{dom}(#1)}
\newcommand{\Fint}{\mathcal{\fint}}
\newcommand{\fformeval}[2]{\widehat{#1}\!\left(#2\right)}

\newcommand{\fnnformeval}[2]{\widehat{\widehat{#1}}\!\left(#2\right)}
\newcommand{\fnformeval}[3]{\fnnformeval{(#1,#2)}{#3}}
\newcommand{\flpcontf}[1]{#1^{\text{\raisebox{0.3ex}{\large$\mkern-1mu\mathord{\cdot}$}}}}

\newcommand{\nfformeval}[2]{\left[#1\right]\!\left(#2\right)}
\newcommand{\nfnnformeval}[2]{\left[\!\left[#1\right]\!\right]\!\left(#2\right)}
\newcommand{\nfnformeval}[3]{\nfnnformeval{(#1,#2)}{#3}}
\newcommand{\flpnotf}{\mathord{\flpcontf{\flpnot}\mkern-2mu}}

\newcommand{\flpnegf}{\mathord{\flpcontf{\flpneg}\mkern-2mu}}

\newcommand{\flpif}[1]{\overset{#1}{\gets}}
\newcommand{\flpnot}{\mathord{\sim}}
\newcommand{\flpneg}{\neg}
\newcommand{\flpand}{\land}

\newcommand{\flpweight}{\vartheta}
\newcommand{\flp}{P}
\newcommand{\flpoperator}[2]{#1_{#2}}
\newcommand{\flpop}[1][\flp]{\flpoperator{T}{#1}}

\newcommand{\flpr}{\atom \flpif{\flpweight}_i \body}

\newcommand{\eflpr}{\literal\lpif{}\body}
\newcommand{\eflprp}{\atom\!\lpif{}\!\body}
\newcommand{\eflprn}{\flpneg\atom\!\lpif{}\!\body}
\newcommand{\body}{\mathcal{B}}
\newcommand{\pbody}{\body^{\not\flpnot}}
\newcommand{\nbody}{\body^{\flpnot}}

\newcommand{\bodies}[1][\literal]{\mathbf{B}_{#1}}

\newcommand{\flptext}{FLP\xspace}
\newcommand{\flpstext}{FLPs\xspace}
\newcommand{\pflptext}{p\flptext}

\newcommand{\nflptext}{n\flptext}
\newcommand{\nflpstext}{n\flpstext}
\newcommand{\eflptext}{e\flptext}
\newcommand{\eflpstext}{e\flpstext}

\newcommand{\ann}[2]{#1\mathord{:}#2}
\newcommand{\gecon}[1]{\mathord{\trianglerighteq_{#1}}}

\newcommand{\Saadop}{O}
\newcommand{\SaadP}{Q}

\newcommand{\Saadredr}[1][\eint]{r_{#1}}
\newcommand{\Saadredp}[1][\eint]{\SaadP_{#1}}

\newcommand{\litTolit}[1][\eint]{\fint_{#1}}

\newcommand{\lpif}{\gets}
\newcommand{\lpnot}{\mathord{\sim}}
\newcommand{\lpand}{\land}

\newcommand{\literals}[1]{#1^{\pm}}

\newcommand{\nliterals}[1]{#1^{-}}
\newcommand{\literal}{\ell}

\newcommand{\modelfor}{\Vdash}

\newcommand{\lset}{V}

\newcommand{\lleq}{\leqslant}

\newcommand{\lub}{\vee}
\newcommand{\glb}{\wedge}
\newcommand{\biglub}{\bigvee\!}
\newcommand{\bigglb}{\bigwedge\!}
\newcommand{\least}{0}
\newcommand{\grtst}{1}

\newcommand{\leastint}{\mathbf{\least}}
\newcommand{\grtstint}{\mathbf{\grtst}}

\newcommand{\fst}[1]{#1_1}
\newcommand{\snd}[1]{#1_2}

\newcommand{\op}{O}
\newcommand{\ap}{\mathcal{A}}

\newcommand{\stable}[1]{#1^{\mathit{st}}}
\newcommand{\stap}{\stable{\ap}}

\newcommand{\lpoperator}[2]{#1_{#2}}

\newcommand{\lpop}[1][\lp]{\lpoperator{T}{#1}}

\newcommand{\lpap}[1][\lp]{\lpoperator{\mathcal{T}}{#1}}

\newcommand{\bil}[1][\lset]{#1^2}

\newcommand{\pleq}{\leq_p}

\newcommand{\plub}{\oplus}
\newcommand{\pglb}{\otimes}

\newcommand{\inttlub}{\sqcup}
\newcommand{\inttglb}{\sqcap}

\newcommand{\biginttlub}{\bigsqcup}
\makeatletter
\providecommand{\bigsqcap}{%
  \mathop{%
    \mathpalette\@updown\bigsqcup
  }%
}
\newcommand*{\@updown}[2]{%
  \rotatebox[origin=c]{180}{$\m@th#1#2$}%
}
\makeatother
\newcommand{\biginttglb}{\bigsqcap}
\newcommand{\tleq}{\leq_t}

\newcommand{\tlub}{\vee_t}
\newcommand{\tglb}{\wedge_t}
\newcommand{\ppleq}{\leqq_p}

\newcommand{\lfp}{\mathit{lfp}}
\newcommand{\gfp}{\mathit{gfp}}

\newcommand{\apKHS}{\ap_{P}} %

\newcommand{\SakamaPFlit}[1][\flp,\sinttup]{\Phi_{#1}}
\newcommand{\SakamaPFlitsup}[1][i]{\Phi_{\flp,\sinttupsup[#1]}}
\newcommand{\SakamaDFlit}[1][\flp,\sinttup]{\Psi_{#1}}
\newcommand{\SakamaDFlitsup}[1][i]{\Psi_{\flp,\sinttupsup[#1]}}
\newcommand{\SakamaWFlit}[1][\flp]{\Theta_{#1}}

\newcommand{\bilToLit}{\zeta}
\newcommand{\teToLit}{\zeta_2}

\newcommand{\PFint}{\sigma}
\newcommand{\PFintsup}[1][i]{\PFint^{(#1)}}
\newcommand{\DFint}{\delta}
\newcommand{\DFintsup}[1][i]{\DFint^{(#1)}}

\newcommand{\litset}{\alpha}
\newcommand{\litsetsup}[2][i]{\litset^{(#1,#2)}}
\newcommand{\litsetalt}{\beta}
\newcommand{\litsetaltsup}[2][i]{\litsetalt^{(#1,#2)}}

\newcommand{\sint}{S}
\newcommand{\sinttup}{\ensuremath{\langle \PFint, \DFint \rangle}}
\newcommand{\sinttupsup}[1][i]{\ensuremath{\langle \PFintsup[#1], \DFintsup[#1] \rangle}}

\newcommand{\comp}[1]{\overline{#1}}

\newcommand{\formula}{F}
\newcommand{\formulaalt}{G}

\longtrue

\begin{document}

\maketitle

\begin{abstract}
  In logic programming, negation can be interpreted in various ways.
Probably best known is the concept of “negation as failure”, where “$\mathit{not}\, p$” is true if we have no evidence for $p$.
On the other hand, strong negation requires that we have evidence for $p$ being false.
Defining semantics for logic programs containing both kinds of negation is a challenging task, and this becomes even more challenging when combining this with other extensions of logic programming, e.g.\ fuzziness.
In this work, we use the framework of approximating fixpoint theory to formulate well-behaved semantics for fuzzy logic programs containing both “by-failure” and strong negation.
We show that this framework generalizes several existing semantics as well as giving rise to a host of new semantics.

\end{abstract}

\section{Introduction}

Logic programs (LPs) are among the oldest rule-based formalisms within the field of artificial intelligence~\cite{KowalskiK71}.
While originally used for declarative programming, their modern incarnation in answer set programming~\cite{BrewkaET11,GebserKKS12} enjoys widespread practical success for declarative problem solving, while using the same basic constructs at the core of the language.
Researchers noted early on that the inherent restriction to two truth values limits applicability, and proposed generalizations from \emph{crisp} (two-valued) to \emph{fuzzy} (many-valued) logics -- for resolution in general~\cite{Lee72}, and for (definite) logic programs in particular \cite{Shapiro83,vanEmden1986quantitative,Hinde86}. %

Naturally, ``weak'' negation (indicating absence of provability and thus also known as negation by failure~\cite{Clark77}) has been combined with fuzziness, leading to \emph{normal} fuzzy logic programs offering various semantics, including stable models~\cite{VanNieuwenborgh2007b,cornejo2018syntax} and well-founded semantics~\cite{LoyerS09}.
On the other hand, crisp normal LPs have also been extended with ``strong'' negation (indicating established falsity~\cite{GelfondL91}), leading to \emph{extended} logic programs with both kinds of negation.
While the original proposal of semantics for extended LPs required answer sets (stable models) of programs to be consistent, that is, disallowed the presence of an atom and its classical negation in the model, it did not take long until \emph{paraconsistent} semantics were proposed for extended LPs -- again both well-founded semantics~\cite{sakama1992extended} and stable model semantics~\cite{SakamaI95}.
The idea of paraconsistent reasoning is that inconsistency is ``allowed'' to occur in a model, but is ``locally contained'' and thus does not lead to the usual \emph{ex falso quodlibet} issues of classical logic~\cite{belnap1977useful}, i.e.\ it avoids trivializing logical reasoning.
And while \emph{extended fuzzy} LPs with stable model semantics do exist, they either do not offer paraconsistent reasoning at all~\cite{Saad09a,JanssenSVC12}, or do not offer it in full generality: some works~\cite{alferes1995logic,damasio1995model,alcantara2005encompassing} base their semantics on the \emph{coherence principle}, which requires them to make restrictive assumptions on the negation functions that can possibly be employed.

The common theme in all of these works on various extensions of normal LPs is their piecemeal nature, the high amount of manual labour involved, and the sheer repetitiveness with which authors took the same ideas, such as the monotone one-step consequence operator of van Emden and Kowalski~\cite{vanEmdenK76} and the reduct of Gelfond and Lifschitz~\cite{GelfondL91}, and manually adapted them to cover the ever next extension.
The result is a scattered landscape with no uniform syntax, semantics with the same name but formally unclear relationships, and varying assumptions on possible sets of truth values and logical connectives.

In this paper, we unify a considerable portion of the existing work in this area.
First, we employ a syntax that covers major recurring features of the literature on extended fuzzy logic programming.
Second, and more importantly, we define an operator that captures a considerable number of existing semantics.
This is possible because we employ the powerful framework of \emph{approximation fixpoint theory} (AFT)~\cite{denecker00approximations}.
AFT reduces the amount of manual labour required to a minimum, as it already incorporates -- within its standard definitions -- ideas that previous authors had to rework explicitly in their respective settings, such as the program reduct of Gelfond and Lifschitz~\cite{GelfondL91}.
More concretely, we show how our \emph{single} operator reconstructs the semantics of Sakama~\cite{sakama1992extended}, Sakama and Inoue~\cite{SakamaI95}, Saad~\cite{Saad09a}, and Cornejo et al.~\cite{Cornejo2020extended}, while also giving rise to new semantics, in particular the first stable model and well-founded semantics offering paraconsistent reasoning for extended fuzzy LPs that allows for non-involutive negations (where double negation is not always identity, e.g.\ Gödel negation).

The rest of the paper proceeds as follows.
In the next section, we give the necessary background.
\Cref{sec:extended-flp} introduces the syntax of the language we use, while \Cref{sec:approximator} defines our main operator.
The following two sections reconstruct existing semantics of extended (fuzzy) logic programs, with \Cref{sec:new-contributions} providing additional results beyond mere reconstruction.
We finally discuss avenues for future work.

\section{Background}

\subsection{Approximation Fixpoint Theory (AFT)}

A particular way of defining semantics for programming languages is to associate a program $P$ with a \emph{transformation} operator $T_P$ that transforms a given input-output relation $R$ into an updated relation $T_P(R)$.
In this view, the semantics of program $P$ is given by the least fixpoint of operator $T_P$.
Van Emden and Kowalski~\cite{vanEmdenK76} applied this way of defining semantics to logic programming, developing the first one-step logical consequence operator and employing its monotonicity in conjunction with \citettarskitheorem to define the semantics of definite logic programs with potentially recursive predicate definitions.

More formally, an \define{operator} is a (total) function \mbox{$\op\colon\lset\to\lset$} on a partially ordered set $(\lset,\lleq)$.
Here, we even require $(\lset,\lleq)$ to be a \define{complete lattice}, that is, we require that every subset \mbox{$S\subseteq\lset$} has a
\define{least upper bound (lub)} \mbox{$\biglub S\in\lset$} and
\define{greatest lower bound (glb)} \mbox{$\bigglb S\in\lset$};
this entails that $(\lset,\lleq)$ has both a least element \mbox{$\least=\bigglb\lset=\biglub\emptyset$} and a greatest element \mbox{$\grtst=\biglub\lset=\bigglb\emptyset$}.
An operator \mbox{$\op\colon\lset\to\lset$} is
\define{monotone} iff \mbox{$x\lleq y$} implies \mbox{$\op(x)\lleq\op(y)$}, and
\define{antimonotone} iff \mbox{$x\lleq y$} implies \mbox{$\op(y)\lleq\op(x)$}.
An element \mbox{$x\in\lset$} with \mbox{$\op(x)=x$} is a \define{fixpoint} of $\op$;
an \mbox{$x\in\lset$} with \mbox{$\op(x)\lleq x$} is a \define{pre-fixpoint} of $\op$;
an \mbox{$x\in\lset$} with \mbox{$x\lleq\op(x)$} is a \define{post-fixpoint} of $\op$.

A fundamental result in lattice theory, \citettarskitheorem, states that whenever $\op$ is a monotone operator on a complete lattice $(\lset,\lleq)$, the set \mbox{$\set{x\in\lset \guard \op(x)=x }$} of its fixpoints forms a complete lattice itself, and so has a least element $\lfp(\op)$ and a greatest element $\gfp(\op)$.
While this result is immensely useful, it is not constructive;
however, constructive versions exist and tell us how to actually construct least fixpoints~\citep{markowsky1976chain,cousot1979constructive}:
To this end, one defines, for any \mbox{$x\in\lset$}, the iterated (transfinite) application of $\op$ to $x$ via
\mbox{$\op^{0}(x)\eqdef x$},
\mbox{$\op^{\ordsucc}(x)\eqdef\op(\op^{\ordn}(x))$} for successor ordinals, and
\mbox{$\op^{\ordlim}(x)\eqdef\biglub\set{\op^{\ordn}(x) \guard \ordn<\ordlim}$} for limit ordinals.
It is then known that (for monotone $\op$) there exists an ordinal $\ordn$ such that \mbox{$\op^{\ordn}(\least)=\lfp(\op)$}~\cite{bourbaki49theoreme};
by duality, we have \mbox{$\gfp(\op)=\op^{\ordnalt}(\grtst)$} for some $\ordnalt$.

In order to apply \citettarskitheorem (or its constructive versions), it is thus necessary to have an operator that is monotone, a condition that is no longer fulfilled when moving from \emph{definite} to \emph{normal} logic programs, i.e., when allowing weak negation in rule bodies.
To still have a useful fixpoint theory in the case of non-monotone operators, Denecker, Marek, and Truszczyński (DMT)~\cite{denecker00approximations} fundamentally generalized the theory underlying the fixpoint-based approach to semantics, thereby founding what is now known as \emph{approximation fixpoint theory}.
Its underlying idea is that when an operator of interest does not have properties that guarantee the existence of fixpoints, one can \emph{approximate} this operator in a more fine-grained algebraic structure where fixpoint existence can be guaranteed.

Formally -- following ideas of Belnap~\cite{belnap1977useful}, Ginsberg~\cite{Ginsberg88a}, and Fitting~\cite{fitting02fixpointsemantics} -- DMT~\cite{denecker00approximations} moved from a complete lattice $(\lset,\lleq)$ to its associated \define{bilattice} on the set \mbox{$\bil\eqdef\lset\times\lset$};
where elements of $\lset$ correspond to interpretations, the {pairs} contained in $\bil$ correspond to \emph{approximations} of such interpretations.
More technically, a pair \mbox{$(x,y)\in\bil$} approximates all \mbox{$z\in\lset$} with \mbox{$x\lleq z\lleq y$}.
A pair \mbox{$(x,y)\in\bil$} is called \define{ordered} iff \mbox{$x\lleq y$}, in other words, iff the interval \mbox{$[x,y]\eqdef\set{z\in\lset\guard x\lleq z\lleq y}$} is non-empty;\footnote{Ordered pairs are called “consistent” in the AFT literature~\citep{denecker00approximations}. In this paper, we will use “consistent” in a logical sense and want to avoid confusion.}
a pair $(x,x)$ is called \define{exact}.

There are two natural orderings on $\bil$:
(1) the \hypertarget{tleq}{\define{truth ordering}} extending the (truth) lattice ordering $\lleq$, \mbox{$(x_1,x_2)\tleq(y_1,y_2)\iffdef x_1\lleq y_1 \tand x_2\lleq y_2$},
and
(2)
the \hypertarget{pleq}{\define{precision ordering}}
\mbox{$(x_1,x_2)\pleq(y_1,y_2)\iffdef x_1\lleq y_1 \tand y_2\lleq x_2$},
comparing intervals by precision of approximation.
The lattice operations induced by $\tleq$ are
\hypertarget{tglb}{glb $\tglb$} with $(x_1,x_2)\tglb(y_1,y_2)\eqdef (x_1\glb y_1,x_2\glb y_2)$ and
\hypertarget{tlub}{lub $\tlub$} with $(x_1,x_2)\tlub(y_1,y_2)\eqdef (x_1\lub y_1,x_2\lub y_2)$.
\begin{longtext}
    The lattice operations induced by $\pleq$ are
    \hypertarget{pglb}{glb $\pglb$} with
    \mbox{$(x_1,y_1)\pglb(x_2,y_2)\eqdef (x_1\glb x_2,y_1\lub y_2)$} and
    \hypertarget{plub}{lub $\plub$} with
    \mbox{$(x_1,y_1)\plub(x_2,y_2)\eqdef (x_1\lub x_2,y_1\glb y_2)$}.
\end{longtext}

Just as pairs of $\bil$ approximate elements of $\lset$, a foundational idea of DMT~\cite{denecker00approximations} was that operators \mbox{$\op\colon\lset\to\lset$} can be approximated by operators on $\bil$:
An \define{approximator} is an operator \mbox{$\ap\colon\bil\to\bil$} that is (a) $\pleq$-monotone and (b) maps exact pairs to exact pairs.
By condition~(b), we say that $\ap$ \define{approximates} the operator \mbox{$\op\colon\lset\to\lset$} with \mbox{$\op(x)=\fst{\ap(x,x)}$} (where \mbox{$\fst{(x,y)}=x$} and \mbox{$\snd{(x,y)}=y$});
by condition~(a), $\ap$ is guaranteed to possess ($\pleq$-least) fixpoints.
An approximator $\ap$ is \define{symmetric} iff for all \mbox{$(x,y)\in\bil$}, we have \mbox{$\snd{\ap(x,y)}=\fst{\ap(y,x)}$};
thus symmetry allows one to specify $\ap$ by giving only $\fst{\ap(\cdot,\cdot)}$.

The main contribution of DMT~\cite{denecker00approximations} was the association of the \define{stable approximator $\stap$} to an approximator $\ap$:
for a complete lattice \mbox{$(\lset,\lleq)$} and an approximator \mbox{$\ap\colon\bil\to\bil$}, define the
\define{stable approximator for $\ap$} as \mbox{$\stap\colon\bil\to\bil$} by
\mbox{$\stap(x,y) \eqdef (\lfp(\fst{\ap(\cdot,y)}),\lfp(\snd{\ap(x,\cdot)}))$}, where $\fst{\ap(\cdot,y)}$ (resp.\ $\snd{\ap(x,\cdot)}$) denotes the operator that maps any $z\in\lset$ to $\fst{\ap(z,y)}$ (resp.\ to $\snd{\ap(x,z)}$).
Among other results, DMT~\cite{denecker00approximations} also showed that this construction is well-defined because
both relevant operators
\mbox{$\fst{\ap(\cdot,y)}\colon\lset\to\lset$} and
\mbox{$\snd{\ap(x,\cdot)}\colon\lset\to\lset$}
are $\lleq$-monotone (as $\ap$ is an approximator) and thus possess least fixpoints.
Furthermore, any approximator maps ordered pairs to ordered pairs, and the stable approximator $\stap$ maps ordered post-fixpoints of $\ap$ to ordered pairs.
Consequently, not only is $\lfp(\ap)$ ordered (and called the \define{Kripke-Kleene fixpoint} of $\ap$), but so is $\lfp(\stap)$, the \define{well-founded fixpoint} of $\ap$.
Moreover, a fixpoint $(x,x)$ of $\stap$, called a \define{stable fixpoint} of $\ap$, always induces a $\lleq$-minimal fixpoint $x$ of $\op$ (when $\ap$ approximates $\op$).
The term “stable” is not coincidental: When using (the normal logic program version of) the transformation operator $T_P$~\cite{vanEmdenK76} as $\op$ and defining a suitable approximator~$\ap$ (see \Cref{sec:approximator} for a generalization of such an approximator), then the exact fixpoints of $\stap$ correspond one-to-one to the stable models of~$P$~\citep{denecker00approximations}.

\subsection{Fuzzy Logic Programming}

Generalizing \emph{crisp} logic programming over \mbox{$\lset=\set{0,1}$}, \emph{fuzzy} logic programming allows more general sets $\lset$ of truth values, with a prominent example being \mbox{$\lset=[0,1]$}, the real unit interval~\citep{Lee72}.
This necessitates also generalizing the syntax of rule bodies from mere conjunctions of atoms as in the crisp case to fuzzy logics' richer offering of connectives~\citep{Hajek1998}.
Formally, given a complete lattice $(\lset,\lleq)$ of truth values,
a set $\Atoms$ of propositional atoms, and
a family $\Flpcon$ of fuzzy connectives (where each $f\in\Flpcon$ has an associated \define{arity} $n\in\N$ denoted by $f^{(n)}$ when important),
designating a finitely representable subset $\lset'\subseteq V$ of truth values that can be explicitly mentioned in programs,%
\iflong\footnote{Allowing truth values to occur explicitly in formulas is with the intended usage of being able to represent \emph{some} truth values in rule bodies.
  This does not necessarily work for arbitrary sets of truth values (e.g.\ it does not work for $\lset=[0,1]\subseteq\R$ by sheer cardinality) -- in those cases we want to allow to restrict the syntax to a meaningful subset of truth values that can be finitely represented, e.g.\ the rational unit interval $\lset'=[0,1]\cap\mathbb{Q}$.}\fi
\ the syntax of a \define{positive fuzzy formula} \eqref{eq:positive-formula} and a \define{positive fuzzy logic program rule} \eqref{eq:weightedflprule} is defined via
\begin{center}
  \begin{minipage}{0.45\textwidth}
    \begin{equation}
      \varphi \ebnfeq c \ebnfalt p \ebnfalt \flpcon(\varphi, \dots ,\varphi)
      \label{eq:positive-formula}
    \end{equation}
  \end{minipage}
  \hfill
  \begin{minipage}{0.45\textwidth}
    \begin{equation}
      p \flpif{\flpweight} \body
      \label{eq:weightedflprule}
    \end{equation}
  \end{minipage}
\end{center}
where $c\in\lset'$, $p\in\Atoms$, and $\flpcon \in \Flpcon$, and where $p\in\Atoms$, $\flpweight\in\lset'$, $\flpif{}$ is an implicator, and $\body$ is a positive fuzzy formula. %
The atom $p$ in \eqref{eq:weightedflprule} is called the \define{head} of the rule, $\body$ is called its \define{body}, and $\flpweight$ its \define{weight} (intuitively indicating the degree of truth assigned to the rule).
A \define{positive fuzzy logic program} (\pflptext) $\flp$ is then a finite set of rules~(\ref{eq:weightedflprule}).

Every connective \mbox{$\flpcon^{(n)}\in\Flpcon$} has an associated truth function \mbox{$\flpcontf{\flpcon}\colon\lset^n\to\lset$} on the lattice $(\lset, \lleq)$ of truth values;
we require that each $\flpcon \in \Flpcon$ that occurs in a rule body is monotone in each argument,
and each implicator $\flpcontf{\flpif{}}$ is monotone in the first and antimonotone in the second argument.

For a complete lattice $(\lset, \lleq)$ of truth values and $\Atoms$ a set of atoms,
a \emph{non-paraconsistent (fuzzy) interpretation} is a total function $\nint \colon \Atoms \to \lset$.
The set of all non-paraconsistent interpretations is denoted by $\Nint$ and again forms a complete lattice with the pointwise lifting of $\lleq$.
The evaluation of atoms by an interpretation $\nint$ extends to arbitrary formulas $\varphi$ via structural induction, where we denote the truth value of $\varphi$ under $\nint$ by $\nfformeval{\nint}{\varphi}$:
for connective \mbox{$f^{(n)}\in\Flpcon$}, we define \mbox{$\nfformeval{\nint}{f(\varphi_1,\ldots,\varphi_n)}\eqdef \flpcontf{f}(\nfformeval{\nint}{\varphi_1},\ldots,\nfformeval{\nint}{\varphi_n})$} (and in the base case clearly $\nfformeval{\nint}{\atom}\eqdef\nint(\atom)$).
Given an interpretation \mbox{$\nint\in\Nint$}, a weighted rule \mbox{$\atom\flpif{\flpweight}\body$} is \emph{satisfied} by $\nint$ iff
\mbox{$\flpweight \lleq \nint(\atom)\flpcontf{\flpif{}} \nfformeval{\nint}{\body}$}.
An interpretation $\nint$ is a \emph{model} of the program $\flp$ iff it satisfies all rules in $\flp$.
Intuitively, $\nint$ satisfies a rule if it makes the rule “at least as true” as its weight $\flpweight$ -- the typical notion of fuzzy modelhood.

To guarantee soundness of inference, fuzzy logic programming usually employs pairs of implicators $\flpif{}$ and conjunctors $\flpand$ that “belong together”:
formally, a pair $(\flpcontf{\flpif{}}, \flpcontf{\flpand})$ is called an \define{adjoint pair}~\cite{MedinaOV01} iff
for all $x,y,z \in \lset$ of a partially ordered set \mbox{$(\lset,\lleq)$, $x \flpcontf{\flpand} y \lleq z \mathrel{\text{ iff }} y \lleq z \flpcontf{\flpif{}} x$}.
In view of adjoint pairs, satisfaction of a rule $\atom\flpif{\flpweight}\body$ by an interpretation $\nint$ can equivalently be stated as \mbox{$\flpweight\flpcontf{\flpand}\nfformeval{\nint}{\body}\lleq\nint(\atom)$}.
For one, this allows one to define a fuzzy generalization of the one-step consequence operator given by van Emden and Kowalski~\cite{vanEmdenK76}:
For \pflptext $\flp$, operator \mbox{$\lpop \colon \Nint \to \Nint$}
\citep{Vojtas01,MedinaOV01}
is defined by setting, for each $\atom\in\Atoms$,
\begin{equation}
  \lpop(\nint)(\atom) \eqdef \biglub \set{\flpweight\flpcontf{\flpand}\nfformeval{\nint}{\body} \guard \mathord{\atom\flpif{\flpweight}\body} \in \flp}
  \label{eq:immediate-consequence-weighted}
\end{equation}
The semantics of a \pflptext $\flp$ can then be characterized by the pre-fixpoints of $\lpop$~\cite[Theorem~2.2]{Vojtas01}:
interpretation \mbox{$\nint\in\Nint$} is a model of $\flp$ iff \mbox{$\lpop(\nint)\lleq\nint$}, generalizing a classical result by van Emden and Kowalski~\cite{vanEmdenK76}.
An equally useful consequence of adjoint pairs is that they allow us to “move” weights of rules into the body, i.e.\ writing \eqref{eq:weightedflprule} equivalently as its \emph{weight-free} version $\atom\flpif{}\flpweight\flpand\body$ (using the $\flpand$ for which $(\flpcontf{\flpif{}},\flpcontf{\flpand})$ is an adjoint pair).
As this allows for a simpler notation without sacrificing expressiveness (e.g.\ the RHS of \eqref{eq:immediate-consequence-weighted} simplifies to \(\biglub \set{\nfformeval{\nint}{\body} \guard \atom\flpif{}\body \in \flp}\)), in what follows we consider only weight-free rules (also for future extensions, i.e.\ normal and extended rules).

\subsection{Normal Fuzzy Logic Programs}
\label{sec:normal-flps}

Syntactically, normal \flpstext (\nflpstext) allow (in their bodies) \define{normal} fuzzy formulas, that are built by extending \eqref{eq:positive-formula} with the alternative $\flpnot p$, where $\flpnot$ is “negation as failure”.
The remaining notions for positive formulas, rules, and programs carry over to the normal case as expected; %
in particular, the evaluation of a normal fuzzy formula $\varphi$ under \mbox{$\nint\in\Nint$} has the additional inductive case \mbox{$\nfformeval{\nint}{\flpnot\varphi}\eqdef\flpnotf\nfformeval{\nint}{\varphi}$}.
This directly extends the one-step consequence operator \eqref{eq:immediate-consequence-weighted} to the case of \emph{normal} \flpstext, with models of $\flp$ and pre-fixpoints of $\lpop$ still being in one-to-one correspondence.
Alas, as in the crisp case, the occurrence of $\flpnot$ leads to $\lpop$ no longer being necessarily monotone, whence several authors generalized (crisp) semantics to the fuzzy case \citep{LoyerS09,JanssenSVC12,cornejo2018syntax}, by hand-crafting the definitions.

Recently, we \cite{KettmannHS25} have presented an approximator $\apKHS$ of $\lpop$ for the case of \nflpstext, and showed that this one approximator (via the standard definitions of approximation fixpoint theory) uniformly reconstructs several hand-crafted semantics:
the well-founded fixpoint of $\apKHS$ coincides with the fuzzy well-founded semantics defined by Loyer and Straccia~\cite{LoyerS09};
the exact stable fixpoints of $\apKHS$ coincide with the fuzzy stable models defined by Cornejo et al.~\cite{cornejo2018syntax}. The operator we will present in Section \ref{sec:approximator} is a generalization of $\apKHS$.
In our previous work \cite{KettmannHS25}, we have not explicitly discussed the work of Janssen et al.~\cite{JanssenSVC12};
however, as the latter base their semantics on the definitions of Medina et al.~\cite{MedinaOV01}, it follows that the semantics used by Janssen et al.~\cite{JanssenSVC12} is also covered by our $\apKHS$~\cite{KettmannHS25}.

\section{Extended Fuzzy Logic Programs}

In this paper, we are interested in \define{extended} fuzzy logic programs.
Firstly, “extended” means that we allow two kinds of negation --
\define{weak (default) negation} $\flpnot$, which captures the absence of a proof, and \define{strong (classical) negation} $\flpneg$, which expresses explicit falsity.
Secondly, we abstract away from concrete sets $\lset$ of truth values, e.g.\ $\lset=\set{\least,\grtst}$ in the Boolean case, or $\lset=[0,1]$ in the case of fuzzy LPs~\cite{Vojtas01}, but merely require the set $\lset$ of truth values to be equipped with a (possibly partial) truth ordering $\lleq$ such that $(\lset,\lleq)$ forms a complete lattice.

Syntactically, the move from normal to \emph{extended} \flpstext sees us replacing atoms with literals across definitions. %
Based on a set $\Atoms$ of propositional atoms, we define the set of \emph{literals} as \mbox{$\literals{\Atoms} \eqdef \Atoms \cup \nliterals{\Atoms}$}, where $\Atoms$ is the set of \emph{positive} literals and \mbox{$\nliterals{\Atoms} \eqdef \set{\flpneg \atom \mid \atom \in \Atoms}$} the set of \emph{negative} literals.
\begin{definition}
  \label{def:syntax}\label{eq:eflprule}
  Let $(\lset,\lleq)$ be a complete lattice of truth values with
  $\lset'\subseteq\lset$ computable and
  $\Atoms$ be a set of atoms.
  With $c\in\lset'$, $\literal\in\literals{\Atoms}$, and $\flpcon\in\Flpcon$, an \define{extended fuzzy formula $\varphi$} is of the form
  \begin{equation}
    \varphi \ebnfeq c \ebnfalt \literal \ebnfalt \flpnot\literal \ebnfalt \flpcon(\varphi, \dots ,\varphi)
    \label{eq:extended-formula}
  \end{equation}
  and an \define{extended fuzzy logic program rule} is of the form $\literal \flpif{} \body$,
  where, differently from \eqref{eq:positive-formula} and \eqref{eq:weightedflprule}, $\literal\in\literals{\Atoms}$, and $\body$ is an extended fuzzy formula.
  An \define{extended fuzzy logic program} (\define{\eflptext}) $\flp$ is then a finite set of (extended fuzzy) rules.
\end{definition}
For the negators $\flpneg$ and $\flpnot$, the corresponding truth functions $\flpnegf$ and $\flpnotf$ must
(1) be antimonotone on $(\lset,\lleq)$; and
(2) behave classically on $\least,\grtst\in\lset$, that is, must satisfy $\flpnegf\least=\grtst$ and $\flpnegf\grtst=\least$ (and equivalently for $\flpnotf$).
Our model theory is motivated by the foundational insights deriving from
(a) paraconsistent semantics for \emph{crisp extended} logic programs (using $\lset=\set{\least,\grtst}$)
\cite{sakama1992extended,SakamaI95}, and
(b) our previous treatment of \emph{fuzzy normal} LPs in approximation fixpoint theory~\cite{KettmannHS25}.
\begin{definition}
  \label{def:interpretation,consistent} \label{def:extended-formula-evaluation}
  A \define{(paraconsistent fuzzy) interpretation} is a {\iflong total\fi} function \mbox{$\fint\colon \Atoms \to \bil$};
  we denote the set of all interpretations by $\Fint$.
  A pair \mbox{$(v_1,v_2)\in\bil$} is \define{consistent} iff \mbox{$v_1\lleq\flpnegf v_2$};
  an \mbox{$\fint\in\Fint$} is \define{consistent} iff $\fint(\atom)$ is consistent for all \mbox{$\atom\in\Atoms$}.

  For an extended fuzzy formula $\varphi$ and a paraconsistent fuzzy interpretation \mbox{$\fint\in\Fint$},
  we inductively define $\fformeval{\fint}{\varphi}$ (the evaluation of $\varphi$ by $\fint$), with base cases
  \mbox{$\fformeval{\fint}{c}\eqdef c$} (for \mbox{$c\in\lset'$}),
  \mbox{$\fformeval{\fint}{\atom}\eqdef\fst{\fint(\atom)}$} and \mbox{$\fformeval{\fint}{\flpneg\atom}\eqdef\snd{\fint(\atom)}$} (for \mbox{$\atom\in\Atoms$}), and
  \mbox{$\fformeval{\fint}{\flpnot\literal}\eqdef\flpnotf\fformeval{\fint}{\literal}$} (for \mbox{$\literal\in\literals{\Atoms}$});
  and \mbox{$\fformeval{\fint}{f(\varphi_1,\ldots,\varphi_n)}\eqdef\flpcontf{f}(\fformeval{\fint}{\varphi_1},\ldots,\fformeval{\fint}{\varphi_n})$} for any \mbox{$f^{(n)}\in\Flpcon$} in the inductive case.
\end{definition}
Intuitively, for interpretation $\fint$ and atom \mbox{$\atom\in\Atoms$}, the value $\fst{\fint(\atom)}$ expresses how \emph{true} $\atom$ is;
similarly, the value $\snd{\fint(\atom)}$ expresses how \emph{false} $\atom$ is.
This representation generalizes Belnap's four-valued logic \cite{belnap1977useful}, which takes into account information about truth and falsity of atoms independently of each other.
A consistent $\fint$ gives an interval $[\fst{\fint(\atom)},\flpnegf\snd{\fint(\atom)}]$ of possible truth values for every $\atom\in\Atoms$.
In the special case of fuzzy logic with \mbox{$\flpnegf x=1-x$}, we obtain that $(x_1,x_2)$ is consistent iff \mbox{$x_1\lleq 1-x_2$} iff \mbox{$x_1+x_2\lleq 1$}.
Note that $\flpnegf$ is only used to determine consistency, but not for evaluating formulas;
also note that \mbox{$\fformeval{\fint}{\varphi}\in\lset$} for any \mbox{$\fint\in\Fint$} and formula $\varphi$:
that is, while interpretations assign \emph{pairs} of truth values to atoms (one for truth and one for falsity), the evaluation of a formula is a \emph{single} truth value.
The truth ordering $\tleq$ on pairs of truth values extends naturally (pointwise) to interpretations:
Formally, for $\fint,\fintalt\in\Fint$ we have
\hypertarget{inttglb}{glb $\inttglb$} with
$\fint\inttglb\fintalt\eqdef\set{\atom\mapsto\fint(\atom)\tglb\fintalt(\atom)\guard\atom\in\Atoms}$, and
\hypertarget{inttlub}{lub $\inttlub$} with
$\fint\inttlub\fintalt\eqdef\set{\atom\mapsto\fint(\atom)\tlub\fintalt(\atom)\guard\atom\in\Atoms}$;
the binary notation carries over to arbitrary subsets $\fintset\subseteq\Fint$, viz.\ $\biginttglb\fintset$ and $\biginttlub\fintset$.

A paraconsistent fuzzy interpretation \mbox{$\fint\in\Fint$} is a \define{model} of an \eflptext rule \mbox{$\literal\flpif{}\body$} iff \mbox{$\fformeval{\fint}{\body}\lleq\fformeval{\fint}{\literal}$}.
$\fint$ is a \define{model} of an \eflptext $\flp$ if it is a model of all rules in $\flp$.

\begin{example}
  \label{exm:toy-example-eflp}
  Consider \mbox{$\lset=\set{\least,\grtst}$} and the two \eflpstext
  \mbox{$\flp_1=\set{ \flpneg q\lpif{} \flpnot p,\; p\lpif{}p }$}
  and
  \mbox{$\flp_2=\set{ \flpneg q\lpif{} \flpneg p,\; p\lpif{}p }$}.
  The $\tleq$-least (and intended) model of $\flp_1$ is interpretation
  $\fint_1=\set{p\mapsto(\least,\least), q\mapsto(\least,\grtst)}$ where $p$ is false (as it cannot be derived in a well-founded way), thus $\flpnot p$ is true and by the first rule $\flpneg q$ is true, while $q$ is false as there is no rule for it.
  In contrast, the least model of $\flp_2$ is given by
  $\fint_2=\set{p\mapsto(\least,\least), q\mapsto(\least,\least)}$ where all literals are false.
\end{example}

The first step in our reconstruction is defining a consequence operator that characterizes this model semantics. %

\begin{definition}
  \label{def:consequence-operator}
  Let $\flp$ be an extended fuzzy logic program.
  Its associated \emph{immediate consequence operator $\lpop$} is given by
  \begin{align*}
    \iflong\lpop(\fint)(\atom) & \eqdef \left(\lpop^+(\fint)(\atom),\lpop^-(\fint)(\atom)\right) \text{ with} \\\fi
    \lpop^+(\fint)(\atom)      & \eqdef \biglub \set{\fformeval{\fint}{\body} \guard \eflprp \in \flp}
    \text{ and }\;
    \lpop^-(\fint)(\atom)  \eqdef \biglub \set{\fformeval{\fint}{\body} \guard \eflprn \in \flp}
  \end{align*}
\end{definition}

$\lpop$ is still a van-Emden-Kowalski $T_P$-operator, i.e.\ exactly characterizes the models through its prefixpoints.
\begin{proposition}
  \label{p:lpop:van-Emden-Kowalski}
  For any \eflptext $\flp$ and \mbox{$\fint\in\Fint$}, %
  $I$ is a model of $\flp$ $\iff$
  \;\mbox{$\lpop(\fint)\tleq\fint$}.
  \begin{longproof}
    We first note that for
    any two paraconsistent fuzzy interpretations $\fint,\fintalt\in\Fint$, we have
    \begin{align*}
      \fint\tleq\fintalt & \iff \forall\atom\in\Atoms: \fst{\fint(\atom)}\lleq\fst{\fintalt(\atom)} \tand \snd{\fint(\atom)}\lleq\snd{\fintalt(\atom)}                                       \\
                         & \iff \forall\atom\in\Atoms: \fformeval{\fint}{\atom}\lleq\fformeval{\fintalt}{\atom} \tand \fformeval{\fint}{\flpneg\atom}\lleq\fformeval{\fintalt}{\flpneg\atom} \\
                         & \iff \forall\literal\in\literals{\Atoms}: \fformeval{\fint}{\literal}\lleq\fformeval{\fintalt}{\literal}
    \end{align*}
    Furthermore, for any rule $\eflprp\in\flp$, we have
    \[
      \fformeval{\fint}{\body} \lleq \lpop^+(\fint)(\atom) = \fst{\lpop(\fint)(\atom)} =  \fformeval{\lpop(\fint)}{\atom}
    \]
    and similarly for any rule $\eflprn\in\flp$, we get
    \[
      \fformeval{\fint}{\body} \lleq \lpop^-(\fint)(\atom) = \snd{\lpop(\fint)(\atom)} = \fformeval{\lpop(\fint)}{\flpneg\atom}
    \]
    whence for any rule $\eflpr\in\flp$, we have
    \begin{gather}
      \tag{$\dagger$}
      \fformeval{\fint}{\body} \lleq \fformeval{\lpop(\fint)}{\literal}
    \end{gather}
    \begin{description}
      \item[\normalfont“if”:]
            Let $\lpop(\fint)\tleq\fint$.
            We directly get
            \[
              \forall\literal\in\literals{\Atoms}: \fformeval{\lpop(\fint)}{\literal}\lleq\fformeval{\fint}{\literal}
            \]
            and with ($\dagger$) obtain that for any $\eflpr\in\flp$,
            \[
              \fformeval{\fint}{\body} \lleq \fformeval{\lpop(\fint)}{\literal} \lleq \fformeval{\fint}{\literal}
            \]
            whence $\fint$ is a model of $\flp$.

      \item[\normalfont“only if”:]
            Assume $\fformeval{\fint}{\body}\lleq\fformeval{\fint}{\literal}$ for all $\eflpr\in\flp$.
            Let $\literal\in\literals{\Atoms}$ be arbitrary and define $\bodies^{\fint}\eqdef\set{ \fformeval{\fint}{\body} \guard \eflpr\in\flp }$.
            By assumption, $\fformeval{\fint}{\literal}$ is an upper bound of $\bodies^{\fint}$.
            Therefore,
            \[
              \fformeval{\lpop(\fint)}{\literal}=\biglub\bodies^{\fint}\lleq\fformeval{\fint}{\literal}
            \]
            Since $\literal$ was chosen arbitrarily, the claim follows.\qed
    \end{description}
  \end{longproof}
\end{proposition}

\begin{longtext}
  We can also show that for formulas not containing default negation $\flpnot$, evaluation of formulas behaves monotonically with respect to the relevant truth orderings.
  For results that are only in the long version of the paper (such as the next one), we use a distinctive numbering system in order to keep the numbers of results in the short version in sync with the long version.

  \begin{longlemma}
    \label{lem:positive-fformeval-p-monotone}
    For all $\fint,\fintalt\in\Fint$ and $\flpnot$-free formula $\varphi$, we have
    \begin{narrowgather}
      \fint\tleq\fintalt \implies \fformeval{\fint}{\varphi}\lleq\fformeval{\fintalt}{\varphi}
    \end{narrowgather}
    \begin{longproof}
      Let $\fint\tleq\fintalt$.
      We use induction on the structure of $\varphi$ to show the claim.
      \begin{description}
        \item[\normalfont$\varphi=\atom$:]
              We have $\fformeval{\fint}{\atom}=\fst{\fint}(\atom)\lleq\fst{\fintalt}(\atom)=\fformeval{\fintalt}{\atom}$.
        \item[\normalfont$\varphi=\flpneg\atom$:]
              We have $\fformeval{\fint}{\flpneg\atom}=\snd{\fint}(\atom)\lleq\snd{\fintalt}(\atom)=\fformeval{\fintalt}{\flpneg\atom}$.
        \item[\normalfont$\varphi=f(\varphi_1,\ldots,\varphi_n)$:]
              By induction hypothesis, for any $1\leq i\leq n$, we have \mbox{$\fformeval{\fint}{\varphi_i}\lleq\fformeval{\fintalt}{\varphi_i}$}.
              The claim now follows from pointwise $\lleq$-monotonicity of $\flpcontf{f}$.\qed
      \end{description}
    \end{longproof}
  \end{longlemma}

  Thus for $\flpnot$-free extended logic programs, the consequence operator is $\tleq$-monotone.
  This will be useful in later reconstructions of stable model semantics, which are based on a reduct program not containing default negation.
\end{longtext}

\begin{shorttext}
  For formulas not containing default negation $\flpnot$, evaluation of formulas behaves monotonically with respect to the relevant truth orderings. This means that  for $\flpnot$-free extended logic programs, the consequence operator is $\tleq$-monotone.
  This will be useful in later reconstructions of stable model semantics, which are based on a reduct program not containing default negation.
\end{shorttext}

\begin{proposition}
  \label{thm:lpop:positive-monotone}
  \iflong
    For a $\flpnot$-free \eflptext $\flp$ and interpretations \mbox{$\fint,\fintalt\in\Fint$}, we have: $\fint\tleq\fintalt\implies\lpop(\fint)\tleq\lpop(\fintalt)$.
  \fi
  \begin{longproof}
    Assume $\fint\tleq\fintalt$.
    Let $\literal\in\literals{\Atoms}$ be arbitrary and denote
    \(
    \bodies \eqdef \set{ \body' \guard \eflpr'\in\flp }
    \)
    as well as
    \[
      \bodies^{\fint}\eqdef\set{\fformeval{\fint}{\body}\guard\body\in\bodies}
      \text{ and }\;
      \bodies^{\fintalt}\eqdef\set{\fformeval{\fintalt}{\body}\guard\body\in\bodies}\!.
    \]
    Now for any $\body\in\bodies$,
    by \Cref{lem:positive-fformeval-p-monotone},
    \(
    \fformeval{\fint}{\body}\lleq\fformeval{\fintalt}{\body}\lleq\biglub\bodies^{\fintalt}
    \)
    whence $\biglub\bodies^{\fintalt}$ is an upper bound of $\bodies^{\fint}$.
    Since $\biglub\bodies^{\fint}$ is the least upper bound of $\bodies^{\fint}$, we obtain
    \[
      \fformeval{\lpop(\fint)}{\literal}=
      \biglub\bodies^{\fint}\lleq\biglub\bodies^{\fintalt}
      =\fformeval{\lpop(\fintalt)}{\literal}
    \]
    which concludes the proof.\qed
  \end{longproof}
\end{proposition}

While this result on $\lpop$ is useful for later reconstructions, it does not provide information on programs with \emph{both} kinds of negation.
There, as in the case of \nflpstext, it is clear that $\lpop$ is not necessarily $\tleq$-monotone and therefore cannot be used to define a (least-fixpoint) semantics.
Approximation fixpoint theory comes to the rescue here:
in the next section, we will define an approximator for $\lpop$ and employ the machinery of AFT to obtain several semantics at once.

\label{sec:extended-flp}

\section{An Approximator for Extended Fuzzy LPs}
\label{sec:approximator}
We now define an approximator of the immediate consequence operator for the full language of \eflpstext with weak negation.
The relevant lattice-theoretic structure for the immediate consequence operator, i.e.\ the operator we approximate, is $(\Fint,\tleq)$.
As is customary for AFT, this means that our approximator will operate on the bilattice based on $(\Fint,\tleq)$, which we denote by $(\bil[\Fint],\ppleq)$ and refer to as the (precision) \define{tetralattice}.\footnote{The notion of tetralattice is applicable since $(\Fint,\tleq)$ is technically isomorphic to the (truth) bilattice based on $(\Nint,\lleq)$.}
More formally, for two pairs $(\lint,\uint),(\lintalt,\uintalt)\in\bil[\Fint]$ we have
$(\lint,\uint)\ppleq(\lintalt,\uintalt)\iffdef\lint\tleq\lintalt\tand\uintalt\tleq\uint$.
The least element of $(\bil[\Fint],\ppleq)$ is the pair $(\leastint,\grtstint)$, where $\leastint\eqdef\set{\atom\mapsto(\least,\least)\guard\atom\in\Atoms}$ and $\grtstint\eqdef\set{\atom\mapsto(\grtst,\grtst)\guard\atom\in\Atoms}$ are the least and greatest element of $(\Fint,\tleq)$, respectively.

\begin{definition}
    \label{def:approximator}
    For \eflptext $\flp$, define symmetric approximator $\lpap\colon\bil[\Fint]\to\bil[\Fint]$ with
    \begin{align*}
        \lpap(\lint, \uint)                & \eqdef \left(\fst{\lpap(\lint,\uint)},\fst{\lpap(\uint,\lint)}\right)                   \\
        \fst{\lpap(\lint, \uint)}(\atom)   & \eqdef \left(\fst{\lpap^+(\lint,\uint)}(\atom),\fst{\lpap^-(\lint,\uint)}(\atom)\right) \\
        \fst{\lpap^+(\lint, \uint)}(\atom) & \eqdef \biglub \set{\fnformeval{\lint}{\uint}{\body} \guard \eflprp \in \flp}           \\
        \fst{\lpap^-(\lint, \uint)}(\atom) & \eqdef \biglub \set{\fnformeval{\lint}{\uint}{\body} \guard \eflprn \in \flp}
    \end{align*}%
    \noindent where by induction %
    $\fnformeval{\lint}{\uint}{\literal}\eqdef\fformeval{\lint}{\literal}$ and
    $\fnformeval{\lint}{\uint}{\flpnot\literal}\eqdef\flpnotf\fformeval{\uint}{\literal}$ for $\literal\in\literals{\Atoms}$, and
    $\fnformeval{\lint}{\uint}{f(\varphi_1,\ldots,\varphi_n)}\eqdef\flpcontf{f}\!\!\left(\fnformeval{\lint}{\uint}{\varphi_1},\ldots,\fnformeval{\lint}{\uint}{\varphi_n}\right)$ for any $f^{(n)}\in\Flpcon$.
\end{definition}
Note that again, $\fnformeval{\lint}{\uint}{\varphi}\in\lset$, so formula evaluation even with elements of the tetralattice yields a single truth value.
We look at an example to illustrate the operator, with special attention to the treatment of weak and strong negation.

\begin{example}
    \label{exm:toy-example-approximator}
    Recall \eflptext $P_1$ (over $\lset=\set{\least,\grtst}$) from Example \ref{exm:toy-example-eflp}.
    Applying the approximator to the pair $(\leastint,\grtstint)$ yields
    $\lpap[{\flp_1}](\leastint,\grtstint)
        =(\fst{\lpap[{\flp_1}](\leastint,\grtstint)},\fst{\lpap[{\flp_1}](\grtstint,\leastint)})
        =(\leastint,\set{p\mapsto(\grtst,\least),q\mapsto(\least,\grtst)})$
    While all literals stay false in the lower bound, in the upper bound we can see that $\flpneg p$ and $q$ turn false because they have no rules.
\end{example}

The operator $\lpap$ is $\ppleq$-monotone, allowing us to generally obtain its least fixpoint by transfinite iteration (i.e.\ $\lfp(\lpap)=(\biginttglb_{i\geq 0}\fst{\lpap^i(\leastint,\grtstint)},\biginttlub_{i\geq 0}\snd{\lpap^i(\leastint,\grtstint)})$).
\begin{longtext}
    \begin{longlemma}
        For any \eflptext $\flp$ and $\lint,\uint,\lintalt,\uintalt\in\Fint$, we have
        \begin{narrowgather}
            (\lint,\uint)\ppleq(\lintalt,\uintalt) \implies \lpap(\lint,\uint)\ppleq\lpap(\lintalt,\uintalt)
        \end{narrowgather}
        \begin{longproof}
            Let $(\lint,\uint)\ppleq(\lintalt,\uintalt)$;
            then $\lint\tleq\lintalt$ and $\uintalt\tleq\uint$.

            We start out with showing a helper claim pertaining to the evaluation of general extended formulas with pairs of interpretations.

            \noindent\begin{myclaim}
                \label{c:fnformeval-monotone}
                For every extended fuzzy formula $\varphi$,
                $(\lint,\uint)\ppleq(\lintalt,\uintalt)$ implies $\fnformeval{\lint}{\uint}{\varphi}\lleq\fnformeval{\lintalt}{\uintalt}{\varphi}$.

                \noindent\begin{claimproof}
                    We use induction on the structure of $\varphi$.
                    \begin{description}
                        \item[\normalfont$\varphi=\literal\in\literals{\Atoms}$:]
                              Then $\fnformeval{\lint}{\uint}{\literal}=\fformeval{\lint}{\literal}\lleq\fformeval{\lintalt}{\literal}=\fnformeval{\lintalt}{\uintalt}{\literal}$ by \Cref{lem:positive-fformeval-p-monotone} as $\lint\pleq\lintalt$.
                        \item[\normalfont$\varphi=\flpnot\literal$:]
                              Then $\fnformeval{\lint}{\uint}{\flpnot\literal}=\flpnotf\fformeval{\uint}{\literal}
                                  \lleq\flpnotf\fformeval{\uintalt}{\literal}=\fnformeval{\lintalt}{\uintalt}{\flpnot\literal}$ where the central inequality is due to \Cref{lem:positive-fformeval-p-monotone} (which yields $\fformeval{\uintalt}{\literal}\lleq\fformeval{\uint}{\literal}$) and $\flpnotf$ being $\lleq$-antimonotone.
                        \item[\normalfont$\varphi=f(\varphi_1,\ldots,\varphi_n)$:]
                              Then by the induction hypothesis, for all $1\leq i\leq n$, we have that $\fnformeval{\lint}{\uint}{\varphi_i}\lleq\fnformeval{\lintalt}{\uintalt}{\varphi_i}$.
                              Overall, we thus get
                              \begin{align*}
                                   & \pheq \fnformeval{\lint}{\uint}{f(\varphi_1,\ldots,\varphi_n)}                                                                                                          \\
                                   & =\flpcontf{f}(\fnformeval{\lint}{\uint}{\varphi_1},\ldots,\fnformeval{\lint}{\uint}{\varphi_n})                 & \text{(Def.~$\fnformeval{\cdot}{\cdot}{\cdot}$)}      \\
                                   & \lleq\flpcontf{f}(\fnformeval{\lintalt}{\uintalt}{\varphi_1},\ldots,\fnformeval{\lintalt}{\uintalt}{\varphi_n}) & \text{(IH and $\flpcontf{f}$ being $\lleq$-monotone)} \\
                                   & =\fnformeval{\lintalt}{\uintalt}{f(\varphi_1,\ldots,\varphi_n)}                                                 & \text{(Def.~$\fnformeval{\cdot}{\cdot}{\cdot}$)}
                              \end{align*}
                              which concludes the proof of the claim.
                    \end{description}
                \end{claimproof}
            \end{myclaim}

            \noindent Now back to the main proof.
            Due to symmetry of $\lpap$, it suffices to show that $\fst{\lpap(\lint,\uint)}(\atom)\tleq\fst{\lpap(\lintalt,\uintalt)}$, that is,
            \[
                \fst{\lpap^+(\lint,\uint)}(\atom) \lleq \fst{\lpap^+(\lintalt,\uintalt)}(\atom)
                \text{ and }
                \fst{\lpap^-(\lint,\uint)}(\atom) \lleq \fst{\lpap^-(\lintalt,\uintalt)}(\atom)
            \]
            which can be shown exactly as in the normal case \cite[Proposition~1]{KettmannHS25}.\qed
        \end{longproof}
    \end{longlemma}
    As $\lpap$ also agrees with $\lpop$ on exact inputs, together with $\ppleq$-monotonicity, this makes $\lpap$ an approximator of $\lpop$.
\end{longtext}
\begin{shorttext}
    As $\lpap$ also agrees with $\lpop$ on exact inputs, $\lpap$ approximates $\lpop$.
\end{shorttext}
\begin{proposition}
    For any \eflptext $\flp$, $\lpap$ is an approximator for $\lpop$.
    \begin{longproof}
        It suffices to show that for any interpretation $\fint\in\Fint$ and formula $\varphi$, we have \mbox{$\fnformeval{\fint}{\fint}{\varphi}=\fformeval{\fint}{\varphi}$}.
        By induction on $\varphi$, we see:
        \[
            \fnformeval{\fint}{\fint}{\literal}=\fformeval{\fint}{\literal} \text{ and }
            \fnformeval{\fint}{\fint}{\flpnot\literal}=\flpnotf\fformeval{\fint}{\literal}=\fformeval{\fint}{\flpnot\literal}\text{ by definition }
        \]
        and for the induction step
        \begin{align*}
            \fnformeval{\fint}{\fint}{f(\varphi_1,\ldots,\varphi_n)}
             & = \flpcontf{f}(\fnformeval{\fint}{\fint}{\varphi_1},\ldots,\fnformeval{\fint}{\fint}{\varphi_n})
            \stackrel{\text{(IH)}}{=}\flpcontf{f}(\fformeval{\fint}{\varphi_1},\ldots,\fformeval{\fint}{\varphi_n}) \\
             & =\fformeval{\fint}{f(\varphi_1,\ldots,\varphi_n)}\qedhere
        \end{align*}
    \end{longproof}
\end{proposition}

\section{Reconstructions: Paraconsistent Semantics for Extended Crisp Logic Programs}
\label{sec:reconstructions-crisp}

In this section, we restrict our attention to the classical two truth values \mbox{$\lset=\set{\least,\grtst}$} and extended rules \mbox{$\literal\lpif{}\body$} with
\(
\body=\literal_1\lpand\ldots\lpand\literal_m \lpand\lpnot\literal_{m+1}\land\ldots\land\lpnot\literal_n
\)
denoting \mbox{$\pbody=\set{\literal_1,\ldots,\literal_m}$} and \mbox{$\nbody=\set{\literal_{m+1},\ldots,\literal_n}$}.
Existing (paraconsistent) semantics for extended logic programs \cite{GelfondL91,sakama1992extended,SakamaI95} typically employ a seemingly different notion of interpretation -- sets \mbox{$S\subseteq\literals{\Atoms}$}.
We start off with showing that our notion of interpretation is actually order-isomorphic to such sets of literals.

\begin{proposition}
  \label{p:bilToLit-order-isomorphism}
  The function
  \( \bilToLit\colon\Fint\to 2^{\literals{\Atoms}} \)
  with
  $\bilToLit(\fint)\eqdef\set{ \atom\in\Atoms \guard \fst{\fint(\atom)}=\grtst} \cup \set{ \flpneg\atom \guard \snd{\fint(\atom)}=\grtst }$
  is an order-isomorphism from $(\Fint,\tleq)$ to $(2^{\literals{\Atoms}},\subseteq)$, that is,
  (1) $\bilToLit$ is a bijection, and
  (2) for all $\fint,\fintalt\in\Fint$, we have $\fint\tleq\fintalt\iff\bilToLit(\fint)\subseteq\bilToLit(\fintalt)$.
  \begin{longproof}
    \begin{enumerate}
      \item[(1)]
            \begin{description}
              \item[\normalfont$\bilToLit$ is injective:]
                    Assume w.l.o.g.\ that $\literal\in\bilToLit(\fint)\setminus\bilToLit(\fintalt)$.
                    Then $\fformeval{\fint}{\literal}\neq\fformeval{\fintalt}{\literal}$ whence $\fint\neq\fintalt$.
              \item[\normalfont$\bilToLit$ is surjective:]
                    We give the inverse of $\bilToLit$.
                    For any \mbox{$S\subseteq\literals{\Atoms}$}, define interpretation $\fint_S$ such that
                    \mbox{$\fint_S(\atom)\eqdef(\card{\set{\atom}\cap S},\card{\set{\flpneg\atom}\cap S})$} for any \mbox{$\atom\in\Atoms$}.
                    It is then easy to see that \mbox{$\bilToLit(\fint_S)=S$}.
            \end{description}
      \item[(2)]
            \begin{description}
              \item[\normalfont“$\implies$”:]
                    Let $\fint\tleq\fintalt$ and consider $\atom\in\bilToLit(\fint)$.
                    Then $\fst{\fint(\atom)}=\grtst$ and by $\fint\tleq\fintalt$ we get $\fst{\fintalt(\atom)}=\grtst$.
                    Thus, $\atom\in\bilToLit(\fintalt)$.
                    Likewise, if $\flpneg\atom\in\bilToLit(\fint)$, then $\snd{\fint(\atom)}=\grtst$.
                    By $\fint\tleq\fintalt$, we get $\snd{\fintalt(\atom)}=\grtst$, whence $\flpneg\atom\in\bilToLit(\fintalt)$.
              \item[\normalfont“$\impliedby$”:]
                    Let $\atom\in\Atoms$ be arbitrary.
                    If $\fst{\fint(\atom)}=\least$ or $\snd{\fint(\atom)}=\least$ there is nothing to show.
                    If $\fst{\fint(\atom)}=\grtst$ then $\atom\in\bilToLit(\fint)\subseteq\bilToLit(\fintalt)$ and thus $\fst{\fintalt(\atom)}=\grtst$.
                    Likewise, if $\snd{\fint(\atom)}=\grtst$ then $\flpneg\atom\in\bilToLit(\fint)\subseteq\bilToLit(\fintalt)$ whence $\snd{\fintalt(\atom)}=\grtst$.
            \end{description}
    \end{enumerate}
  \end{longproof}
\end{proposition}

This order-isomorphism thus preserves all lubs and glbs: for example, we have $\bilToLit(\biginttlub\fintset)=\bigcup\bilToLit(\fintset)$ for any $\fintset\subseteq\Fint$.
We will employ this later on.

\subsection{Paraconsistent Well-Founded Semantics \citep{sakama1992extended}}

Sakama~\cite{sakama1992extended} defines a well-founded semantics for extended (crisp) logic programs.
His definition uses operators and is based on \citeauthor{Przymusinski}{przymusinski89}'s definition \citeyearpar{przymusinski89} of the well-founded semantics for normal logic programs~\citep{GelderRS91}.
Again, \citeauthor{Sakama}{sakama1992extended}'s model theory is seemingly different:
an interpretation \mbox{$\sinttup\in 2^{\literals{\Atoms}}\times 2^{\literals{\Atoms}}$} is a pair of sets of literals where $\PFint$ represents \emph{proven facts} while $\DFint$ represents \emph{default facts}.
Similarly to (simple) paraconsistent interpretations, for \mbox{$\lset= \set{0,1}$} these interpretations are isomorphic to pairs \mbox{$(\lint,\uint)\in\bil[\Fint]$} of paraconsistent interpretations via
\mbox{\(\teToLit\colon\bil[\Fint] \to 2^{\literals{\Atoms}}\times 2^{\literals{\Atoms}}\)} with
\mbox{$\teToLit(\lint,\uint) \eqdef \left(\bilToLit(\lint), \comp{\bilToLit(\uint)}\right)$}, where
\mbox{$\comp{\sint}\eqdef\literals{\Atoms}\setminus\sint$}. %

For an interpretation $\sinttup$, the \define{satisfaction} of a formula $\formula$, denoted $\sinttup\modelfor\formula$, is defined inductively as follows:
for any literal $\literal\in\literals{\Atoms}$,
$\sinttup\modelfor \literal$ iff $\literal\in\PFint$, and
$\sinttup\modelfor \lpnot\literal$ iff $\literal\in\DFint$; then,
$\sinttup\modelfor \formula\lpand\formulaalt$ iff $\sinttup\modelfor\formula$ and $\sinttup\modelfor\formulaalt$.

\citeauthor{Sakama}{sakama1992extended}'s paraconsistent well-founded semantics~\citeyearpar{sakama1992extended} for extended LPs is then defined using the following operators:
\begin{align*}
  \SakamaPFlit(\litset)       & \eqdef \set{ \literal\in\literals{\Atoms}\guard \exists\literal\lpif\body\in\flp: \pbody\subseteq \PFint\cup\litset \text{ and } \nbody\subseteq\DFint }                 \\
  \SakamaDFlit(\litset)       & \eqdef \set{ \literal\in\literals{\Atoms}\guard \forall\literal\lpif\body\in\flp: \pbody\cap(\litset\cup\DFint)\neq\emptyset \text{ or } \nbody\cap\PFint\neq\emptyset } \\
  \SakamaWFlit(\PFint,\DFint) & \eqdef (\PFint\cup\lfp(\SakamaPFlit),\DFint\cup\gfp(\SakamaDFlit))
\end{align*}
where the \define{Sakama well-founded model} of $\flp$ is given by $\lfp(\SakamaWFlit)$ in the complete lattice $(2^{\literals{\Atoms}}\times2^{\literals{\Atoms}},\leq_S)$ with $(A,B)\leq_S (C,D)$ iff $A\subseteq C$ and $B\subseteq D$.

In what follows, we will formally establish that the Sakama well-founded model of an extended \flptext $\flp$ coincides (modulo $\teToLit$) with the standard AFT definition of the well-founded fixpoint of our approximator $\lpap$ from \Cref{def:approximator}.%
\iflong~
To this end, we first introduce some more notation.
To indicate the interpretations during the construction of the Sakama well-founded model, we use superscripts, that is,
\mbox{$(\PFintsup[0],\DFintsup[0])\eqdef (\emptyset,\emptyset)$} and subsequently \mbox{$(\PFintsup[i+1],\DFintsup[i+1]) \eqdef \SakamaWFlit(\PFintsup,\DFintsup)$} for $i\geq 0$.
Analogously, we denote the steps of the corresponding nested fixpoint constructions of newly proven facts at step $i$ by
$\litsetsup{0}\eqdef\emptyset$ and $\litsetsup{j+1}\eqdef\SakamaPFlitsup(\litsetsup{j})$, and
for the default facts use $\litsetaltsup{0}\eqdef\literals{\Atoms}$ and $\litsetaltsup{j+1}\eqdef\SakamaDFlitsup(\litsetaltsup{j})$.
Similarly, for the construction of the well-founded fixpoint of $\lpap$, we denote the sequence of constructed interpretation pairs via
$(\lintsup[0],\uintsup[0])\eqdef(\leastint,\grtstint)$ and
$(\lintsup[i+1],\uintsup[i+1])\eqdef (\lfp(\fst{\lpap(\cdot, \uintsup)}),\lfp(\fst{\lpap(\cdot, \lintsup)}))$ for $i\geq 0$.
For the nested least fixpoint constructions, we denote
$\fintsup{0}\eqdef\leastint$ and $\fintsup{j+1} \eqdef \fst{\lpap(\fintsup{j},\uintsup)}$ when constructing new lower bound $\lintsup[i+1]$, and
$\fintaltsup{0}\eqdef\leastint$ and $\fintaltsup{j+1} \eqdef \fst{\lpap(\fintaltsup{j},\lintsup)}$ for new upper bound $\uintsup[i+1]$.

\begin{longtext}

  We first observe that any rule body $\body$ evaluates to $\grtst$ under a pair of paraconsistent interpretations $(\fint,\fintalt)$ iff
  $\body$ is satisfied by the corresponding Sakama interpretation $\teToLit(\fint,\fintalt)$.
  \begin{longlemma}
    \label{lem:sakama-wf:bodies}
    For any $\fint,\fintalt\in\Fint$ and rule body $\body$, we have
    \[
      \fnformeval{\fint}{\fintalt}{\body}=\grtst \iff \tuple{\bilToLit(\fint),\comp{\bilToLit(\fintalt)}} \modelfor \body .
    \]
    \begin{longproof}
      For all interpretations $\fint,\fintalt\in\Fint$ and body formula $\body = \literal_1\lpand\ldots\lpand\literal_m\lpand\lpnot\literal_{m+1}\land\ldots\land\lpnot\literal_n$, we have
      \begin{align*}
         & \phiff \rtuple{\bilToLit(\fint),\comp{\bilToLit(\fintalt)}}\modelfor\body                                                                                                                                                                                                             \\
         & \iff \pbody\subseteq\bilToLit(\fint) \text{ and } \nbody\subseteq\comp{\bilToLit(\fintalt)}                                                                                                                                                                                           \\
         & \iff \text{ for all } \literal\in\pbody: \fformeval{\fint}{\literal} = \grtst \text{ and }\text{for all } \literal\in\nbody: \fformeval{\fintalt}{\literal} = \least                                                                                                                  \\
         & \iff \fformeval{\fint}{\literal_1}\flpcontf{\lpand}\ldots\flpcontf{\lpand}\fformeval{\fint}{\literal_m} = \grtst \text{ and } \flpcontf{\lpnot}\fformeval{\fintalt}{\literal_{m+1}}\flpcontf{\lpand}\ldots\flpcontf{\lpand}\flpcontf{\lpnot}\fformeval{\fintalt}{\literal_n} = \grtst \\
         & \iff \fnformeval{\fint}{\fintalt}{\body}=\grtst\qedhere
      \end{align*}
    \end{longproof}
  \end{longlemma}

  In turn, satisfaction of a rule body coincides with the truth of the rule head in the approximator's return value.

  \begin{longlemma}
    \label{lem:sakama-wf:literals}
    For all $\fint,\fintalt\in\Fint$ and literals $\literal\in\literals{\Atoms}$, we have
    \begin{narrowgather}
      \fformeval{\fst{\lpap(\fint,\fintalt)}}{\literal}=\grtst
      \iff
      \exists\literal\flpif{}\body\in\flp \text{ with } \fnformeval{\fint}{\fintalt}{\body}=\grtst
    \end{narrowgather}
    \begin{longproof}
      We do a case distinction on the syntactic form of $\literal$:
      \begin{description}
        \item[\normalfont$\literal=\atom\in\Atoms$:]
              From \Cref{def:approximator}, it follows directly that
              \begin{align*}
                 & \phiff \fformeval{\fst{\lpap(\fint,\fintalt)}}{\atom}=\grtst                                                  \\
                 & \iff \fst{\fst{\lpap(\fint,\fintalt)}(\atom)}=\grtst                                                          \\
                 & \iff \fst{\lpap^+(\fint,\fintalt)}(\atom)=\grtst                                                              \\
                 & \iff \biglub\set{ \fnformeval{\fint}{\fintalt}{\body} \guard \atom\lpif{}\body\in\flp }=\grtst                \\
                 & \iff \exists \text{ rule } \atom\flpif{}\body\in\flp \text{ with } \fnformeval{\fint}{\fintalt}{\body}=\grtst
              \end{align*}
        \item[\normalfont$\literal=\flpneg\atom$ for some $\atom\in\Atoms$:]
              Likewise, \Cref{def:approximator} yields
              \begin{align*}
                 & \phiff \fformeval{\fst{\lpap(\fint,\fintalt)}}{\flpneg\atom}=\grtst                                                          \\
                 & \iff \snd{\fst{\lpap(\fint,\fintalt)}(\atom)}=\grtst                                                                         \\
                 & \iff \fst{\lpap^-(\fint,\fintalt)}(\atom)=\grtst                                                                             \\
                 & \iff \biglub\set{ \fnformeval{\fint}{\fintalt}{\body} \guard \flpneg\atom\lpif{}\body\in\flp }=\grtst                        \\
                 & \iff \exists \text{ rule } \flpneg\atom\flpif{}\body\in\flp \text{ with } \fnformeval{\fint}{\fintalt}{\body}=\grtst\qedhere
              \end{align*}
      \end{description}
    \end{longproof}
  \end{longlemma}

  This leads to the first correspondence result on the involved operators:
  under certain conditions, $\lpap$ and $\SakamaPFlit$ yield the same results (modulo isomorphism).
\end{longtext}

\begin{shorttext}
  Under certain conditions, $\lpap$ and $\SakamaPFlit$ yield the same results (modulo isomorphism).
\end{shorttext}

\begin{proposition}
  \label{sakama-wf:fst}
  For all interpretations $\lintsup,\uintsup,\fintsup{j}\in\Fint$ with
  $\lintsup\tleq\fintsup{j}$,
  and for all literals $\literal\in\literals{\Atoms}$,
  denoting
  $\bilToLit(\fintsup{j})=\litsetsup{j}$ and
  $\teToLit(\lintsup,\uintsup)=(\PFintsup,\DFintsup)$,
  we have
  \(
  \fformeval{\fst{\lpap(\fintsup{j},\uintsup)}}{\literal} = \grtst \iff \literal\in\SakamaPFlitsup(\litsetsup{j})
  \).
  \begin{longproof}
    From condition $\lintsup\tleq\fintsup{j}$, we obtain directly that $\fintsup{j}\tlub\lintsup=\fintsup{j}$,
    thus
    \[
      \tag{$\dagger$}
      \litsetsup{j}\cup\PFintsup=\bilToLit(\fintsup{j})\cup\bilToLit(\lintsup)
      =\ \bilToLit(\fintsup{j}\tlub\lintsup)=\bilToLit(\fintsup{j})=\litsetsup{j}
    \]

    Now:
    \begin{align*}
       & \phiff \fformeval{\fst{\lpap(\fintsup{j},\uintsup)}}{\literal} = \grtst                                                                                                                \\
       & \iff \exists\literal\flpif{}\body\in\flp \text{ with } \fnformeval{\fintsup{j}}{\uintsup}{\body}=\grtst                                & \qquad \text{(\Cref{lem:sakama-wf:literals})} \\
       & \iff \exists\literal\flpif{}\body\in\flp \text{ with } \tuple{\bilToLit(\fintsup{j}),\comp{\bilToLit(\uintsup)}} \modelfor \body       & \qquad \text{(\Cref{lem:sakama-wf:bodies})}   \\
       & \iff \exists\literal\flpif{}\body\in\flp \text{ with } \tuple{\litsetsup{j},\DFintsup} \modelfor \body                                 & \qquad \text{(Notation)}                      \\
       & \iff \exists\literal\flpif{}\body\in\flp \text{ with } \tuple{\litsetsup{j}\cup\PFintsup,\DFintsup} \modelfor \body                    & \qquad \text{($\dagger$)}                     \\
       & \iff \exists\literal\flpif{}\body\in\flp \text{ with } \pbody\subseteq\litsetsup{j}\cup\PFintsup \text{ and } \nbody\subseteq\DFintsup & \qquad \text{(Def.~$\modelfor$)}              \\
       & \iff \literal\in\SakamaPFlitsup(\litsetsup{j})\qedhere
    \end{align*}
  \end{longproof}
\end{proposition}

\begin{longtext}
  Since $(\lintsup,\uintsup)_{i\geq 0}$ is a chain, these “certain conditions” will eventually hold.

  \begin{longlemma}
    \label{lem:sakama-wf:exists-lower}
    For all \mbox{$i\geq 0$} there is a \mbox{$j^*\geq 0$} such that for all \mbox{$k\geq j^*$},  \mbox{$\lintsup\tleq\fintsup{k}$}.

    \begin{longproof}
      Set $\fintsup{j^*}\eqdef\lintsup[i+1]=\lfp(\fst{\lpap(\cdot,\uintsup)})$.
      It remains to show that for all $k\geq j^*$, we have $\lintsup\tleq\fintsup{k}$.
      For this, we first show that \mbox{$(\lintsup,\uintsup)_{i\geq 0}$} is a $\ppleq$-chain:
      Since \mbox{$(\lintsup[0],\uintsup[0])=(\leastint,\grtstint)$} is the least element of \mbox{$(\bil[\Fint],\ppleq)$}, it is clear that \mbox{$(\lintsup[0],\uintsup[0])\ppleq(\lintsup[1],\uintsup[1])$}.
      It follows from Proposition~20 of \cite{denecker00approximations} that $\stable{\lpap}$ is $\ppleq$-monotone and therefore
      $(\lintsup[1],\uintsup[1])=\stable{\lpap}(\lintsup[0],\uintsup[0])\ppleq\stable{\lpap}(\lintsup[1],\uintsup[1])=(\lintsup[2],\uintsup[2])$.
      Thus, we can show by (transfinite) induction that for all $i\geq 0$, we have $(\lintsup,\uintsup)\ppleq(\lintsup[i+1],\uintsup[i+1])$ and $(\lintsup,\uintsup)_{i\geq 0}$ is a $\ppleq$-chain.
      (In particular, $(\lintsup)_{i\geq 0}$ is a $\tleq$-chain.)
      By an argument similar to the one above, it follows that $(\fintsup{j})_{j\geq 0}$ is a $\tleq$-chain.
      Then for any $k\geq j^*$,
      \[
        \lintsup
        \tleq \lintsup[i+1]
        =  \fintsup{j^*}
        \tleq \fintsup{k}\qedhere
      \]
    \end{longproof}
  \end{longlemma}

  This allows to show correspondence along the iteration sequence $\fintsup{0}, \fintsup{1}, \ldots$ for constructing a new lower bound for the given upper bound $\uintsup$.
  \begin{longlemma}
    \label{lem:sakama-wf:lower}
    For all \mbox{$i,j\geq 0$},
    we have
    \mbox{$\bilToLit(\fintsup{j})=\litsetsup{j}$}
    whenever
    \mbox{$\lintsup\tleq\fintsup{j}$}.
    \begin{longproof}
      The base case $j=0$ holds by definition, as we have $\fintsup{0}=\leastint=\litsetsup{0}$.
      For the induction step,
      \begin{align*}
        \bilToLit(\fintsup{j+1})
         & = \bilToLit(\fst{\lpap(\fintsup{j},\uintsup)})                                                                                               \\
         & = \set{ \literal\in\literals{\Atoms} \guard \fformeval{\fst{\lpap(\fintsup{j},\uintsup)}}{\literal}=\grtst }                                 \\
         & = \SakamaPFlitsup(\litsetsup{j})                                                                             & \text{(\Cref{sakama-wf:fst})} \\
         & = \litsetsup{j+1}\qedhere
      \end{align*}
    \end{longproof}
  \end{longlemma}
\end{longtext}
As the sequences for constructing the next lower bound (true literals) coincide, also their least fixpoints coincide.

\begin{proposition}
  \label{cor:sakama-wf:fst-same-fixpoints}
  For any $(\lintsup,\uintsup)$, $\bilToLit(\lfp(\fst{\lpap(\cdot, \uintsup)}))=\lfp(\SakamaPFlitsup)$.
  \begin{longproof}
    \begin{align*}
      \bilToLit(\lfp(\fst{\lpap(\cdot, \uintsup)}))
       & = \bilToLit\!\left( \biginttlub_{j\geq 0}\fintsup{j} \right)                                                \\
       & = \bigcup_{j\geq 0}\bilToLit(\fintsup{j})                    & \text{(\Cref{p:bilToLit-order-isomorphism})} \\
       & = \bigcup_{j\geq 0}\litsetsup{j}                             & \text{(\Cref{lem:sakama-wf:lower})}          \\
       & = \lfp(\SakamaPFlitsup)\qedhere
    \end{align*}
  \end{longproof}
\end{proposition}

A similar, but “doubly negated” version exists for the second components.
We begin with a technical helper for dealing with Sakama's $\SakamaDFlit$ operator.

\begin{proposition}
  \label{sakama-wf:snd-operator}
  For all $\PFint,\DFint,\litsetalt\subseteq\literals{\Atoms}$ and $\literal\in\literals{\Atoms}$,
  \begin{narrowgather}
    \literal\notin\SakamaDFlit(\litsetalt) \iff
    \exists \literal\lpif\body\in\flp: (\comp{\litsetalt\cup\DFint},\comp{\PFint})\modelfor\body
  \end{narrowgather}
  \begin{longproof}
    Using $A\subseteq B \iff A\cap\comp{B}=\emptyset$
    for any $A,B\subseteq\literals{\Atoms}$,
    \begin{align*}
       & \phiff \literal\notin\SakamaDFlit(\litsetalt)                                                                                                \\
       & \iff \text{not } (\forall \literal\lpif\body\in\flp: \pbody\cap(\litsetalt\cup\DFint)\neq\emptyset \text{ or }\nbody\cap\PFint\neq\emptyset) \\
       & \iff \exists\literal\lpif\body\in\flp: \pbody\cap(\litsetalt\cup\DFint)=\emptyset \text{ and } \nbody\cap\PFint=\emptyset                    \\
       & \iff \exists\literal\lpif\body\in\flp: \pbody\subseteq\comp{\litsetalt\cup\DFint} \text{ and } \nbody\subseteq\comp{\PFint}                  \\
       & \iff \exists\literal\lpif\body\in\flp: (\comp{\litsetalt\cup\DFint},\comp{\PFint})\modelfor\body\qedhere
    \end{align*}
  \end{longproof}
\end{proposition}

This can be used to show (unconditional) correspondence of the operators for upper bounds and default facts.
\begin{longtext}
  \begin{longlemma}
    \label{lem:sakama-wf:snd}
    For all interpretations $\lintsup,\uintsup,\fintsup{j}\in\Fint$
    and for all literals $\literal\in\literals{\Atoms}$,
    denoting
    $\comp{\bilToLit(\fintaltsup{j})}=\litsetaltsup{j}$ and
    $\teToLit(\lintsup,\uintsup)=(\bilToLit(\lintsup),\comp{\bilToLit(\uintsup)})=(\PFintsup,\DFintsup)$,
    we have
    \(
    \fformeval{\fst{\lpap(\fintaltsup{j},\lintsup)}}{\literal} = \grtst \iff \literal\notin\SakamaDFlitsup(\litsetaltsup{j})
    \).
    \begin{longproof}
      We first establish that for all $i,j\geq 0$, we have $\fintaltsup{j}\tleq\uintsup$:
      $(\fintaltsup{j})_{j\geq 0}$ is a chain with least upper bound $\biginttlub_{j\geq 0}\fintaltsup{j}=\lfp(\fst{\lpap(\cdot,\lintsup)})=\uintsup[i+1]$.
      Now for any $j\geq 0$, we have
      \mbox{$\fintaltsup{j}\tleq\uintsup[i+1]\tleq\uintsup$}
      where the rightmost inequality holds because $(\lintsup,\uintsup)_{i\geq 0}$ is a chain.

      Now $\fintaltsup{j}\tleq\uintsup$ implies $\bilToLit(\fintaltsup{j})\subseteq\bilToLit(\uintsup)$ and thus $\comp{\bilToLit(\uintsup)}\subseteq\comp{\bilToLit(\fintaltsup{j})}$,
      therefore we obtain:
      \begin{gather}
        \tag{$\ddagger$}
        \litsetaltsup{j}\cup\DFintsup =\comp{\bilToLit(\fintaltsup{j})}\cup\comp{\bilToLit(\uintsup)} =\comp{\bilToLit(\fintaltsup{j})} =\litsetaltsup{j}
      \end{gather}
      Consequently,
      \begin{align*}
         & \phiff \fformeval{\fst{\lpap(\fintaltsup{j},\lintsup)}}{\literal} = \grtst                                                                                             \\
         & \iff \exists\literal\lpif{}\body\in\flp: \fnformeval{\fintaltsup{j}}{\lintsup}{\body}=\grtst                                                                           \\
         & \iff \exists\literal\lpif{}\body\in\flp: \rtuple{\bilToLit(\fintaltsup{j}),\comp{\bilToLit(\lintsup)}}\modelfor\body  & \text{(\Cref{lem:sakama-wf:bodies})}           \\
         & \iff \exists\literal\lpif{}\body\in\flp: \rtuple{\comp{\litsetaltsup{j}},\comp{\PFintsup}}\modelfor\body              & \text{(Notation)}                              \\
         & \iff \exists\literal\lpif{}\body\in\flp: \rtuple{\comp{\litsetaltsup{j}\cup\DFintsup},\comp{\PFintsup}}\modelfor\body & \text{($\ddagger$)}                            \\
         & \iff \literal\notin\SakamaDFlitsup(\litsetaltsup{j})                                                                  & \text{(\Cref{sakama-wf:snd-operator})}\qedhere
      \end{align*}
    \end{longproof}
  \end{longlemma}

  Thus, we obtain a correspondence along the sequence $\fintaltsup{0}, \fintaltsup{1}, \ldots$ for constructing $\uintsup[i+1]$ given $\lintsup$.
  \begin{longlemma}
    \label{lem:sakama-wf:upper}
    For all \mbox{$i,j\geq 0$} we have
    \mbox{$\comp{\bilToLit(\fintaltsup{j})}=\litsetaltsup{j}$}.
    \begin{longproof}
      We prove by induction.
      The base case $j=0$ holds by definition, as $\comp{\bilToLit(\fintsup{0})}=\comp{\bilToLit(\leastint)}=\comp{\emptyset}=\literals{\Atoms}=\litsetaltsup{0}$.
      For the induction step,
      \begin{align*}
        \comp{\bilToLit(\fintaltsup{j+1})} & = \comp{\bilToLit(\fst{\lpap(\fintaltsup{j},\lintsup)})}                                                                                                  \\
                                           & = \comp{\set{ \literal\in\literals{\Atoms} \guard \fformeval{\fst{\lpap(\fintaltsup{j},\lintsup)}}{\literal}=\grtst}}                                     \\
                                           & = \SakamaDFlitsup(\litsetaltsup{j})                                                                                   & \text{(\Cref{lem:sakama-wf:snd})} \\
                                           & = \litsetaltsup{j+1}\qedhere
      \end{align*}
    \end{longproof}
  \end{longlemma}
\end{longtext}
Thus, again, the sequences for constructing a new upper bound (default facts) coincide, and so do their least fixpoints.

\begin{proposition}
  \label{cor:sakama-wf:snd-same-fixpoints}
  For any $(\lintsup,\uintsup)$, $\comp{\bilToLit(\lfp(\fst{\lpap(\cdot, \lintsup)}))}=\gfp(\SakamaDFlitsup)$.
  \begin{longproof}
    \begin{align*}
      \comp{\bilToLit(\lfp(\fst{\lpap(\cdot, \lintsup)}))}
       & = \comp{\bilToLit\!\left(\biginttlub_{j\geq 0}\fintaltsup{j}\right)}                                                \\
       & = \comp{\bigcup_{j\geq 0}\bilToLit(\fintaltsup{j})}                  & \text{(\Cref{p:bilToLit-order-isomorphism})} \\
       & = \comp{\bigcup_{j\geq 0}\litsetaltsup{j}}                           & \text{(\Cref{lem:sakama-wf:upper})}          \\
       & = \bigcap_{j\geq 0}\litsetaltsup{j}                                  & \text{(de Morgan)}                           \\
       & = \gfp(\SakamaDFlitsup)\qedhere
    \end{align*}
  \end{longproof}
\end{proposition}

\begin{longtext}
  \noindent Thus, we work our way up the well-founded induction chain.

  \begin{longlemma}
    \label{lem:sakama-wf:chain}
    For all $i\geq 0$, we have
    \(
    \teToLit(\lintsup,\uintsup) = (\PFintsup,\DFintsup)
    \).
    \begin{longproof}
      The induction base (case $i=0$) holds by definition\/:
      \begin{multline*}
        \teToLit(\lintsup[0],\uintsup[0])
        =\teToLit(\leastint,\grtstint)
        =\left(\bilToLit(\leastint),\comp{\bilToLit(\grtstint)}\right)=
        \left(\emptyset,\comp{\literals{\Atoms}}\right)
        =(\emptyset,\emptyset)
        =(\PFintsup[0],\DFintsup[0])
      \end{multline*}
      In the induction step, assume the induction hypothesis.
      Now:
      \begin{align*}
                                                                      & \pheq \teToLit(\lintsup[i+1],
        \uintsup[i+1])                                                                                                                                    \\
                                                                      & = \teToLit(\lintsup\inttlub\lintsup[i+1],
        \uintsup\inttglb\uintsup[i+1])                                                                                                                    \\
                                                                      & = \left(\bilToLit(\lintsup\inttlub\lintsup[i+1]),
        \comp{\bilToLit(\uintsup\inttglb\uintsup[i+1])}\right)                                                                                            \\
                                                                      & = \left(\bilToLit(\lintsup)\cup\bilToLit(\lintsup[i+1]),
        \comp{\bilToLit(\uintsup)\cap\bilToLit(\uintsup[i+1])}\right) & \text{(\Cref{p:bilToLit-order-isomorphism})}                                      \\
                                                                      & = \left(\bilToLit(\lintsup)\cup\bilToLit(\lintsup[i+1]),
        \comp{\bilToLit(\uintsup)}\cup\comp{\bilToLit(\uintsup[i+1])}\right)                                                                              \\
                                                                      & = \left(\bilToLit(\lintsup)\cup\bilToLit(\lfp(\fst{\lpap(\cdot, \uintsup)})),
        \comp{\bilToLit(\uintsup)}\cup\comp{\bilToLit(\lfp(\fst{\lpap(\cdot, \lintsup)}))}\right)                                                         \\
                                                                      & = \left(\bilToLit(\lintsup)\cup\lfp(\SakamaPFlitsup),
        \comp{\bilToLit(\uintsup)}\cup\gfp(\SakamaDFlitsup)\right)    & \text{(\Cref{cor:sakama-wf:fst-same-fixpoints,cor:sakama-wf:snd-same-fixpoints})} \\
                                                                      & = (\PFintsup\cup\lfp(\SakamaPFlitsup),
        \DFintsup\cup\gfp(\SakamaDFlitsup))                           & \text{(IH)}                                                                       \\
                                                                      & = (\PFintsup[i+1],\DFintsup[i+1])\qedhere
      \end{align*}
    \end{longproof}
  \end{longlemma}
\end{longtext}

As the two sequences of constructed interpretations are essentially the same, it thus follows that their least fixpoints also coincide (modulo $\teToLit$, respectively).
\fi

\begin{theorem}
  \label{thm:sakama-wf:main}
  For any extended LP $\flp$, we have $\teToLit(\lfp(\stable{\lpap}))=\lfp(\SakamaWFlit)$.
  \begin{longproof}
    We have the following identities\/:
    \begin{align*}
       & \pheq \teToLit(\lfp(\stable{\lpap}))                                                                                                                                             \\
       & = \teToLit\!\left(\biginttlub_{i\geq 0}\lintsup,\biginttglb_{i\geq 0}\uintsup\right)                                                                                             \\
       & = \left(\bilToLit\!\left(\biginttlub_{i\geq 0}\lintsup\right),\comp{\bilToLit\!\left(\biginttglb_{i\geq 0}\uintsup\right)}\right) & \text{(Def.~$\teToLit$)}                     \\
       & = \left(\bigcup_{i\geq 0}\bilToLit(\lintsup),\comp{\bigcap_{i\geq 0}\bilToLit(\uintsup)}\right)                                   & \text{(\Cref{p:bilToLit-order-isomorphism})} \\
       & = \left(\bigcup_{i\geq 0}\bilToLit(\lintsup),\bigcup_{i\geq 0}\comp{\bilToLit(\uintsup)}\right)                                   & \text{(de Morgan)}                           \\
       & = \left(\bigcup_{i\geq 0}\PFintsup,\bigcup_{i\geq 0}\DFintsup\right)                                                              & \text{(\Cref{lem:sakama-wf:chain})}          \\
       & = \lfp(\SakamaWFlit)\qedhere
    \end{align*}
  \end{longproof}
\end{theorem}

\subsection{Paraconsistent Stable Models \citep{SakamaI95}}
\label{sec:sakama-inoue}

While Sakama and Inoue~\cite{SakamaI95} define their semantics for \emph{disjunctive} extended logic programs (i.e.\ allowing for disjunction in the head), we consider only single-literal heads here and leave the disjunctive case for future work.
The paraconsistent stable model semantics for extended LPs by Sakama and Inoue~\cite{SakamaI95} is defined on the basis of a reduct:
Given a program $\flp$ and an interpretation $\fint$, the \define{reduct} of $\flp$ w.r.t.\ $\fint$ is the program
$\flp^{\fint}$ containing exactly all rules $\literal\lpif{}\pbody$ for which there is a rule $\literal\lpif{}\body\in\flp$ with $\fformeval{\fint}{\literal'}=\least$ for all $\literal'\in\nbody$.
The reduct $\flp^{\fint}$ does not contain default negation $\flpnot$, whence $\flpop[{\flp^{\fint}}]$ is $\tleq$-monotone by \Cref{thm:lpop:positive-monotone} and thus possesses a least fixpoint that corresponds to the unique least model of $\flp^{\fint}$~(\Cref{p:lpop:van-Emden-Kowalski}).
An interpretation $\fint$ is then a \define{stable model} of $\flp$ iff $\fint=\lfp(\flpop[{\flp^{\fint}}])$~\cite{SakamaI95}.

These considerations suggest a clear path for showing equivalence of \citeauthor{Sakama and Inoue}{SakamaI95}'s stable models \citeyearpar{SakamaI95} with the stable fixpoints of $\lpap$ as obtained by AFT.
The first step is showing that the operator for revising the lower bound equals the \citeauthor{van Emden and Kowalski}{vanEmdenK76} operator of the reduct.

\begin{proposition}
    \label{p:sakama-inoue:operator-reduct}
    For any extended LP $\flp$ and $\fint\in\Fint$, we have $\flpop[{\flp^{\fint}}]=\fst{\lpap(\cdot,\fint)}$.
    \begin{longproof}
        Let $\fintalt\in\Fint$ and $\atom\in\Atoms$ be arbitrary.
        We have
        \begin{align*}
             & \pheq \flpop[{\flp^{\fint}}](\fintalt)(\atom)                                                                                                                                                                                                                                   \\
             & = \left( \biglub\set{ \fformeval{\fintalt}{\body'} \guard \atom\lpif{}\body' \in \flp^{\fint} },\biglub\set{ \fformeval{\fintalt}{\body'} \guard \flpneg\atom\lpif{}\body' \in \flp^{\fint} } \right)                                                                           \\
             & = \left( \biglub\set{ \fformeval{\fintalt}{\pbody} \guard \atom\lpif{}\body \in \flp \text{ and } \fformeval{\fint}{\nbody}=\grtst }, \biglub\set{ \fformeval{\fintalt}{\pbody} \guard \flpneg\atom\lpif{}\body \in \flp \text{ and } \fformeval{\fint}{\nbody}=\grtst} \right) \\
             & = \left( \biglub\set{ \fnformeval{\fintalt}{\fint}{\body} \guard \atom\lpif{}\body \in \flp }, \biglub\set{ \fnformeval{\fintalt}{\fint}{\body} \guard \flpneg\atom\lpif{}\body \in \flp } \right)                                                                              \\
             & = \fst{\lpap(\fintalt,\fint)}(\atom)\qedhere
        \end{align*}
    \end{longproof}
\end{proposition}

It thus follows that the least fixpoints of these operators coincide and thus also the semantics defined through those.

\begin{theorem}
    \label{thm:sakama-inoue:equivalence}
    Let $\flp$ be an extended logic program and $\fint\in\Fint$.
    $\fint$ is a \citeauthor{Sakama and Inoue}{SakamaI95}~\citeyearpar{SakamaI95} stable model of $\flp$ if and only if $(\fint,\fint)$ is a stable fixpoint of $\lpap$.
    \begin{longproof}
        The following are equivalent:
        (1) $\fint$ is a stable model of $\flp$;
        (2) \mbox{$\fint=\lfp(\flpop[{\flp^{\fint}}])$};
        (3) \mbox{$\fint=\lfp(\fst{\lpap(\cdot,\fint)})$};
        (4) $(\fint,\fint)$ is a stable fixpoint of $\lpap$.\qed
    \end{longproof}
\end{theorem}

As Sakama and Inoue~\cite{SakamaI95} generalize the semantics of Gelfond and Lifschitz~\cite{GelfondL91}, it is clear that for the consistent case, our approximator (and AFT-based stable model semantics) also covers the (non-paraconsistent) stable model semantics of Gelfond and Lifschitz~\cite{GelfondL91}.

\begin{corollary}
    \label{c:gelfond-lifschitz:equivalence}
    Let $\flp$ be a crisp extended LP.
    A consistent interpretation $\fint$ is a \citeauthor{Gelfond and Lifschitz}{GelfondL91} \citeyearpar{GelfondL91} stable model of $\flp$ iff $(\fint,\fint)$ is a stable fixpoint of $\lpap$.
\end{corollary}

\section{Reconstructions: Stable-Model Semantics for Extended Fuzzy Logic Programs}
\label{sec:reconstructions-fuzzy}

We now again consider the general case of truth values $\lset$, that is, no longer restrict to the two-valued case.
This in particular subsumes the case of multi-adjoint fuzzy logic programming, where we reconstruct two stable model semantics for extended \emph{fuzzy} logic programs.
It should be noted that neither of the two existing approaches we subsequently reconstruct allows for \emph{paraconsistent} fuzzy reasoning, so we effectively add this capability to these approaches.

\subsection{Saad's Stable Models for Extended Fuzzy Logic Programs \citep{Saad09a}}
\label{sec:saad-answer-set}

Saad~\cite{Saad09a} uses a slightly different syntax: instead of the rules themselves, all \emph{literals} occurring in rules are annotated.
Concretely, a \citeauthor{Saad}{Saad09a}~rule \citeyearpar{Saad09a} has the form
\begin{equation}
  \label{eq:saad-rule}
  \ann{\literal_0}{c_0} \lpif{} \ann{\literal_1}{c_1}, \ldots, \ann{\literal_m}{c_m}, \flpnot\ann{\literal_{m+1}}{c_{m+1}}, \ldots, \flpnot\ann{\literal_n}{c_n}
\end{equation}
\noindent%
with \mbox{$\literal_i\in\literals{\Atoms}$} and \mbox{$c_i\in\lset'$} for \mbox{$0\leq i\leq n$}, where
intuitively $\ann{\literal}{c}$ expresses a lower truth value bound $c$ for literal $\literal$.
The relevant notion of interpretation for the work of Saad~\cite{Saad09a} is the set $\Eint$ of \emph{extended} fuzzy interpretations, where $\eint\in\Eint$ is a \emph{partial} function \mbox{$\eint\colon\literals{\Atoms}\to [0,1]$}.
By permitting literals to remain \emph{undefined}, the interpretation can distinguish between explicit falsity and the absence of information.
Then, $\eint$ \define{satisfies} $\ann{\literal}{c}$ iff \mbox{$\eint(\literal)\geq c$} and $\literal$ is in the domain of $\eint$ (denoted $\literal\in\dom$),
$\eint$ \define{satisfies} $\flpnot\ann{\literal}{c}$ iff $\literal\notin\dom$ or [$\literal\in\dom$ and \mbox{$\eint(\literal)<c$}],
and $\eint$ \define{satisfies} a rule \eqref{eq:saad-rule} iff the head is satisfied or some body literal is not satisfied.
An interpretation $\eint$ is considered \define{inconsistent} if there exist complementary literals $\literal,\flpneg\literal\in\dom$ such that $\eint(\flpneg\literal)\neq 1-\eint(\literal)$.

As in crisp logic programs, \emph{fuzzy stable models} are defined in terms of a reduct.
Given a \citeauthor{Saad}{Saad09a} program $\SaadP$ and a (guessed) interpretation $\eint$, the \define{reduct} w.r.t. $\eint$ is defined as \mbox{$\Saadredp \eqdef \set{\Saadredr\mid r\in\SaadP \text{ and } \eint \text{ satisfies all } \flpnot\ann{\literal_j}{c_j} \text{ in } r}$}, where $\Saadredr$ is obtained from $r$ by removing all negative body literals. %
Then, Saad \cite{Saad09a} shows that $\eint\in\Eint$ is a \define{fuzzy stable model} of $\SaadP$ iff it is the least fixpoint of the immediate consequence-operator $\Saadop_{\Saadredp}$ on the complete lattice $(\Eint,\leq_A)$ with $\eintalt_1\leq_A\eintalt_2$ iff $\dom[\eintalt_1]\subseteq\dom[\eintalt_2]$ and $\eintalt_1(\literal)\leq\eintalt_2(\literal)$ for all $\literal\in\literals{\Atoms}$.
For a \citeauthor{Saad}{Saad09a} program $\SaadP$, the operator $\Saadop_{\SaadP} \colon \Eint\to\Eint$ is defined by:
\[
  \Saadop_{\SaadP}(\eint)(\literal) \eqdef \max\set{c\guard
    \begin{array}{l}
      \ann{\literal}{c} \lpif{}
      \ann{\literal_1}{c_1}, \ldots, \ann{\literal_m}{c_m},
      \flpnot\ann{\literal_{m+1}}{c_{m+1}}, \ldots, \flpnot\ann{\literal_n}{c_n} \in \SaadP \\
      \text{and } \eint \text{ satisfies all body literals of the rule}
    \end{array}}
\]
and is undefined otherwise (i.e., whenever $\ell$ has no satisfied rules in $\SaadP$).
However, the least fixpoint of $\Saadop_{\SaadP}$ may be inconsistent.
Following the approach of Gelfond and Lifschitz~\cite{GelfondL91}, Saad~\cite{Saad09a} stipulates that any inconsistent fuzzy stable model is replaced by the mapping $\literals{\Atoms}\to\set{\grtst}$.
In what follows, we reconstruct this least-fixpoint characterization within our framework (by means of a translation).

Firstly, to an extended interpretation $\eint\in\Eint$, we associate a corresponding paraconsistent interpretation \mbox{$\litTolit\colon\Atoms\cup\set{d_{\literal}\guard\literal\in\literals{\Atoms}}\to\bil$}, where the fresh atoms $d_{\literal}$ serve to record whether a literal $\literal$ belongs to the domain of $\eint$.
We set
\mbox{$\fst{\litTolit(\atom)} \eqdef \eint(\atom)$} if \mbox{$\atom\in\dom$} and $\least$ otherwise;
\mbox{$\snd{\litTolit(\atom)} \eqdef \eint(\flpneg\atom)$} if \mbox{$\flpneg\atom\in\dom$} and $\least$ otherwise, for $\atom\in\Atoms$;
\mbox{$\litTolit(d_{\literal})\eqdef(1,0)$} if \mbox{$\literal\in\dom$} and $(0,0)$ otherwise.

The idea for our translation is to introduce, for each \mbox{$c\in\lset'$} occurring in a \citeauthor{Saad}{Saad09a} program \citeyearpar{Saad09a}, a new unary connective $\gecon{c}$ with associated truth function
$\flpcontf{\gecon{c}}(x)\eqdef \grtst \text{ if } x\geq c \text{ and } \least \text{ otherwise}$, and to use it to check lower bounds in rule bodies.
For any literal $\literal$ occurring as \mbox{$\ann{\literal}{0}$}, the treatment is different:
$\gecon{0}(\literal)$ simply becomes the atom $d_{\literal}$ used to check whether “$\literal\in\dom$”.
\begin{definition}
  \label{def:saad:translation:rule}
  A \citeauthor{Saad}{Saad09a} rule \eqref{eq:saad-rule} is translated into \eflptext rule
  \mbox{$\literal_0\flpif{}c_0\flpand\pbody\flpand\nbody$} with
  \mbox{$\pbody\eqdef \gecon{c_1}(\literal_1)\flpand \ldots \flpand \gecon{c_m}(\literal_m)$} and
  \mbox{$\nbody\eqdef \flpnot \gecon{c_{m+1}}(\literal_{m+1}) \flpand \ldots \flpand\flpnot \gecon{c_n}(\literal_n)$}.
\end{definition}

\begin{example}
  Consider $\SaadP=\set{\ann{p}{1} \lpif{} \flpnot\ann{q}{0}}$.
  The rule is satisfied by $\eint= \set{p\mapsto 1}$ and $\eintalt=\set{q\mapsto 0}$ with $\eint$ being the only fuzzy stable model.
  The only stable model of the translation $\SaadP'= \set{p\flpif{} \grtst\flpand\flpnot d_q}$ is $\fint_\eint=\set{p\mapsto (1,0);q,d_q\mapsto(0,0)}$.
\end{example}

First, we show that this translation preserves rule body satisfaction.

\begin{lemma}
  \label{lem:saadop-body-equal}
  Let $r$ be a Saad rule \eqref{eq:saad-rule}.
  For any extended interpretation $\eint\in\Eint$ with corresponding paraconsistent interpretation $\litTolit$, we have
  \[
    \eint \text{ satisfies the body of } r \iff \fformeval{\litTolit}{c_0\flpand \pbody \flpand \nbody} \geq c_0,
  \]
  where $c_0 \flpand \pbody \flpand \nbody$ is the body of the translation of $r$.
  \begin{longproof}
    Let $r = \ann{\literal}{c_0} \flpif{} \ann{\literal_1}{c_1}, \ldots, \ann{\literal_m}{c_m}, \flpnot\ann{\literal_{m+1}}{c_{m+1}}, \ldots, \flpnot\ann{\literal_n}{c_n}$.
    By definition, $\eint$ satisfies the body of $r$ iff all body literals are satisfied. We separately consider the positive and negative parts of the rule body.

    \noindent For a positive body literal $\ann{\literal}{c}$, we have
    \begin{align*}
       & \phiff \eint \text{ satisfies } \ann{\literal}{c}                   \\
       & \iff \eint(\literal) \geq c                                         \\
       & \iff \fformeval{\litTolit}{\literal}\geq c                          \\
       & \iff \flpcontf{\gecon{c}}(\fformeval{\litTolit}{\literal}) = \grtst \\
       & \iff \fformeval{\litTolit}{\gecon{c}(\literal)} = \grtst
    \end{align*}
    whence we obtain that all positive body literals of $r$ are satisfied
    iff $\flpcontf{\gecon{c}}(\fformeval{\litTolit}{\literal}) = \grtst$ for all $1\leq i\leq m$
    iff $\fformeval{\litTolit}{\pbody} = \grtst$.

    For a negative body literal $\flpnot\ann{\literal}{c}$, we do a case distinction on whether $c>0$.
    \begin{enumerate}
      \item $c>0$.
            \begin{align*}
               & \phiff \eint \text{ satisfies } \flpnot\ann{\literal}{c}                                                                                         \\
               & \iff  \literal \notin \dom \text{ or } [\literal\in\dom \text{ and } \eint(\literal) < c]                                                        \\
               & \iff  \fformeval{\litTolit}{d_{\literal}}=0 \text{ or } [\fformeval{\litTolit}{d_{\literal}}=1 \text{ and } \fformeval{\litTolit}{\literal} < c] \\
               & \iff  \fformeval{\litTolit}{d_{\literal}}=0 \text{ or } \fformeval{\litTolit}{\literal} < c                                                      \\
               & \iff  \fformeval{\litTolit}{\literal}<c                                                                                                          \\
               & \iff  \flpcontf{\gecon{c}}(\fformeval{\litTolit}{\literal}) = \least                                                                             \\
               & \iff  \flpnotf\flpcontf{\gecon{c}}(\fformeval{\litTolit}{\literal}) = \grtst                                                                     \\
               & \iff  \fformeval{\litTolit}{\flpnot\gecon{c}(\literal)} = \grtst
            \end{align*}
      \item $c=0$.
            \begin{align*}
               & \phiff \eint \text{ satisfies } \flpnot\ann{\literal}{0}                                  \\
               & \iff  \literal \notin \dom \text{ or } [\literal\in\dom \text{ and } \eint(\literal) < 0] \\
               & \iff  \literal \notin \dom                                                                \\
               & \iff  \litTolit(d_\literal)=0                                                             \\
               & \iff  \flpnotf\litTolit(d_\literal) = \grtst                                              \\
               & \iff  \fformeval{\litTolit}{\flpnot d_{\literal}}=\grtst                                  \\
               & \iff  \fformeval{\litTolit}{\flpnot\gecon{c}(\literal)}=\grtst
            \end{align*}
    \end{enumerate}
    Thus all negative body literals of $r$ are satisfied
    iff $\fformeval{\litTolit}{\flpnot\gecon{c_j}(\literal_j)}=1$ for all $m+1\leq j\leq n$
    iff $\fformeval{\litTolit}{\nbody} = 1$.

    Hence, $\eint$ satisfies the body of $r$ iff $\fformeval{\litTolit}{\body} = \fformeval{\litTolit}{c_0 \flpand \pbody \flpand \nbody} \geq c_0$. \qed

  \end{longproof}
\end{lemma}

\begin{definition}
  \label{def:saad:translation:program}
  A \citeauthor{Saad}{Saad09a} program $\SaadP$ is translated into an extended fuzzy logic program $\flp_{\SaadP}$ (\Cref{def:syntax}) rule by rule;
  whenever for some $\literal\in\literals{\Atoms}$ both $\ann{\literal}{0}$ and $\ann{\literal}{c}$ for some $c>0$ occur in $\SaadP$, we add a rule $d_{\literal}\flpif{}\gecon{c}(\literal)$ to $\flp_{\SaadP}$ (for each such $c$).
\end{definition}

The rules  $d_{\literal}\flpif{}\gecon{c}(\literal)$ in the translation are necessary to avoid translated programs allowing additional (stable) models.
E.g., consider
$Q = \set{ \ann{p}{1}\flpif{}\flpnot\ann{p}{0} }$
with “translation”
$Q' = \set{ p\lpif{}1\flpand \flpnot d_p}$
having $\set{p\mapsto(1,0),d_p\mapsto(0,0)}$ as stable fixpoint of $\lpap[Q']$, although $Q$ does not have a stable model.
In any case, we have the following correspondence between the consequence operators $\Saadop_{\SaadP}$ and $\lpop[\SaadP']$.

\begin{lemma}
  \label{lem:saadop-lpop}
  Let $\SaadP$ be an extended fuzzy program according to \cite{Saad09a}, $\SaadP'$ its translation as in \Cref{def:saad:translation:program}, and $\eint\in\Eint$ an extended fuzzy interpretation with counterpart $\litTolit$.
  Then for all $\literal\in\dom[\Saadop_\SaadP(\eint)]$, we have $\fformeval{\lpop[\SaadP'](\litTolit)}{\literal}=\Saadop_\SaadP(\eint)(\literal)$.
  \begin{longproof}
    We do a case distinction on the syntactic form of $\literal$.

    Let \mbox{$\literal=\atom\in\dom[\Saadop_\SaadP(\eint)]$}.
    From \Cref{def:consequence-operator}, it follows that
    \begin{align*}
       & \pheq \fformeval{\lpop[\SaadP'](\litTolit)}{\atom}
      = \fst{\lpop[\SaadP'](\litTolit)(\atom)} = \lpop[{\SaadP'}]^+(\litTolit)(\atom)                      \\
       & = \biglub \left\{\fformeval{\litTolit}{\body} \;\middle|\; \atom\lpif{}\body \in \SaadP' \right\} \\
       & = \max\set{c\guard
        \begin{array}{l}
          \ann{\literal}{c} \lpif{}
          \ann{\literal_1}{c_1}, \ldots, \ann{\literal_m}{c_m},                                 \\
          \flpnot\ann{\literal_{m+1}}{c_{m+1}}, \ldots, \flpnot\ann{\literal_n}{c_n} \in \SaadP \\
          \text{ and } \eint \text{ satisfies all body literals}
        \end{array}}
       & \text{(\Cref{lem:saadop-body-equal})}                                                             \\
       & = \Saadop_{\SaadP}(\eint)(\atom)
       & \text{(Definition of $\Saadop_\SaadP$)}
    \end{align*}
    Now let $\literal=\neg\atom\in\dom[\Saadop_\SaadP(\eint)]$.
    Likewise, \Cref{def:consequence-operator} yields
    \begin{align*}
       & \pheq \fformeval{\lpop[\SaadP'](\litTolit)}{\neg\atom}
      = \snd{\lpop[\SaadP'](\litTolit)(\atom)} = \lpop[{\SaadP'}]^-(\litTolit)(\atom)                          \\
       & = \biglub \left\{\fformeval{\litTolit}{\body} \;\middle|\; \neg\atom\lpif{}\body \in \SaadP' \right\} \\
       & = \max\set{c\guard
        \begin{array}{l}
          \ann{\literal}{c} \lpif{}
          \ann{\literal_1}{c_1}, \ldots, \ann{\literal_m}{c_m},                                 \\
          \flpnot\ann{\literal_{m+1}}{c_{m+1}}, \ldots, \flpnot\ann{\literal_n}{c_n} \in \SaadP \\
          \text{and } \eint \text{ satisfies all body literals}
        \end{array}}
       & \text{(\Cref{lem:saadop-body-equal})}                                                                 \\
       & = \Saadop_{\SaadP}(\eint)(\neg\atom)
       & \text{(Definition of $\Saadop_\SaadP$)}\qedhere
    \end{align*}
  \end{longproof}
\end{lemma}

\noindent This culminates in the desired correspondence on the relevant (reduct) operators (\Cref{lem:saad-extended:operators-equal}) and hence this section's main result, the correspondence of the semantics on programs (\Cref{thm:saad:main}).

\begin{lemma}
  \label{lem:saad-extended:operators-equal}
  Let $\SaadP$ be a \citeauthor{Saad}{Saad09a}~program \citeyearpar{Saad09a}, $\SaadP'$ its translation as in \Cref{def:saad:translation:program}, and $\eint\in\Eint$ an extended fuzzy interpretation with counterpart $\litTolit$.
  For any extended fuzzy interpretation $\eintalt\in\Eint$ with counterpart $\litTolit[\eintalt]$, we have
  \[
    \fformeval{\fst{\lpap[\SaadP'](\litTolit[\eintalt],\litTolit)}}{\literal}=\Saadop_{\Saadredp}(\eintalt)(\literal) \text{ if } \literal\in\dom[\Saadop_{\Saadredp}(\eintalt)]
  \]
  \begin{longproof}
    We do a case distinction on the syntactic form of $\literal$.

    Let $\literal=\atom\in\dom[\Saadop_{\Saadredp}(\eintalt)]$.
    From \Cref{def:approximator}, it follows that
    \begin{align*}
       & \pheq \fformeval{\fst{\lpap[\SaadP'](\litTolit[\eintalt],\litTolit)}}{\atom}
      = \fst{\fst{\lpap[{\SaadP'}](\litTolit[\eintalt],\litTolit)}(\atom)}                                                                                                                                    \\
       & = \fst{\lpap[{\SaadP'}]^+(\litTolit[\eintalt],\litTolit)}(\atom)                                                                                                                                     \\
       & = \biglub \set{\fnformeval{\litTolit[\eintalt]}{\litTolit}{\body} \guard \atom\lpif{}\body \in \SaadP'}
       & \text{(\Cref{def:approximator})}                                                                                                                                                                     \\
       & = \biglub \set{c_0 \flpcontf{\flpand}\fformeval{\litTolit[\eintalt]}{\pbody}\flpcontf{\flpand} \fformeval{\litTolit}{\nbody} \guard \atom\lpif{}\body \in \SaadP'}
       & \text{(Def.)}                                                                                                                                                                                        \\
       & = \biglub \set{c_0 \flpcontf{\flpand}\fformeval{\litTolit[\eintalt]}{\pbody} \guard p\lpif{}\body \in \Saadredp'}
       & \text{(Def. $\Saadredp$)}                                                                                                                                                                            \\
       & = \biglub \set{\fformeval{\litTolit[\eintalt]}{\body} \guard p\lpif{}\body \in \Saadredp'}
       & \text{(Def. $\Saadredp$)}                                                                                                                                                                            \\
       & = \lpop[\Saadredp']^+(\litTolit[\eintalt])(\atom)
       & \text{(Def.)}                                                                                                                                                                                        \\
       & = \Saadop_{\Saadredp}(\eintalt)(\atom)                                                                                                                             & \text{(\Cref{lem:saadop-lpop})}
    \end{align*}

    Let $\literal=\neg\atom\in\dom[\Saadop_{\Saadredp}(\eintalt)]$.
    Likewise, \Cref{def:approximator} yields
    \begin{align*}
       & \pheq \fformeval{\fst{\lpap[\SaadP'](\litTolit[\eintalt],\litTolit)}}{\neg\atom}
      = \snd{\fst{\lpap[{\SaadP'}](\litTolit[\eintalt],\litTolit)}(\atom)}                                                                                                               \\
       & = \fst{\lpap[{\SaadP'}]^-(\litTolit[\eintalt],\litTolit)}(\atom)                                                                                                                \\
       & = \biglub \set{\fnformeval{\litTolit[\eintalt]}{\litTolit}{\body} \guard p\lpif{}\body \in \SaadP'}
       & \text{(\Cref{def:approximator})}                                                                                                                                                \\
       & = \biglub \set{c_0 \flpcontf{\flpand}\fformeval{\litTolit[\eintalt]}{\pbody}\flpcontf{\flpand} \fformeval{\litTolit}{\nbody} \guard \neg\atom\lpif{}\body \in \SaadP'}
       & \text{(Def.)}                                                                                                                                                                   \\
       & = \biglub \set{c_0 \flpcontf{\flpand}\fformeval{\litTolit[\eintalt]}{\pbody} \guard \neg\atom\lpif{}\body \in \Saadredp'}
       & \text{(Def. $\Saadredp$)}                                                                                                                                                       \\
       & = \biglub \set{\fformeval{\litTolit[\eintalt]}{\body} \guard \neg\atom\lpif{}\body \in \Saadredp'}
       & \text{(Def. $\Saadredp$)}                                                                                                                                                       \\
       & = \lpop[\Saadredp']^-(\litTolit[\eintalt])(\atom)
       & \text{(Def.)}                                                                                                                                                                   \\
       & = \Saadop_{\Saadredp}(\eintalt)(\neg \atom)
       & \text{(\Cref{lem:saadop-lpop})}                                                                                                                                        \qedhere
    \end{align*}
  \end{longproof}
\end{lemma}

\begin{theorem}
  \label{thm:saad:main}
  Consider an extended fuzzy LP $\SaadP$ according to Saad~\cite{Saad09a} and its translation $\SaadP'$ obtained via \Cref{def:saad:translation:program}.
  An interpretation $\eint\in\Eint$ is a Saad~\cite{Saad09a} fuzzy stable model of $\SaadP$ iff $(\litTolit,\litTolit)$ is a stable fixpoint of $\lpap[\SaadP']$.
  \begin{longproof}
    The proof is similar to the proof of \Cref{thm:sakama-inoue:equivalence}:
    \begin{align*}
       & \phantom{\text{iff }} \eint \text{ is a fuzzy answer set of } \SaadP                                                                                         \\
       & \text{iff } \eint \text{ is the least model of } \Saadredp                                      & \text{(Def.)}                                              \\
       & \text{iff } \eint \text{ is the least fixpoint of } \Saadop_{\Saadredp}                         & \text{(Def.)}                                              \\
       & \text{iff } \litTolit \text{ is the least fixpoint of } \fst{\lpap[{\SaadP'}](\cdot,\litTolit)} & \text{(\Cref{lem:saad-extended:operators-equal})}          \\
       & \text{iff } (\litTolit,\litTolit) \text{ is a stable fixpoint of } \lpap[{\SaadP'}]             & \text{(Def.)}                                     \qedhere
    \end{align*}
  \end{longproof}
\end{theorem}

\subsection{Stable Models for Extended Fuzzy LPs by Cornejo et al.~\cite{Cornejo2020extended}}
\label{sec:cornejo-stable}

Generalizing earlier work on \emph{normal} logic programs~\cite{cornejo2018syntax}, Cornejo et al.~\cite{Cornejo2020extended} define a (non-paraconsistent) stable model semantics for \emph{extended} (multi-adjoint) \flpstext.
In this section, we recall their semantics and show how we can map their syntax into ours to obtain a one-to-one-correspondence of the respective stable models.

In their syntax for extended fuzzy rules~\cite{Cornejo2020extended}, they do not have explicit classical negation, but instead allow a special kind of $n$-ary aggregation connective $f_e$ such that for some $m$, \mbox{$0\leq m\leq n$}, the function $\flpcontf{f_e}$ is $\lleq$-monotone in arguments \mbox{$1,\ldots,m$} and $\lleq$-antimonotone in arguments \mbox{$m+1,\ldots,n$},\iflong\footnote{In other words, they consider a fuzzy generalization of \emph{bipolar} Boolean functions \cite{BaumannS17}.}\fi\xspace where the underlying idea however still is that such extended aggregators can be expressed by body formulas similar to \eqref{eq:extended-formula}, employing monotone connectives on top of negation $\flpnot$.
In contrast to the classical literature on extended logic programs, Cornejo et al.~\cite{Cornejo2020extended} have no negative literals in rule heads, as there is no classical negation symbol in the syntax.
Instead, they allow \emph{constraints}, rules whose head is a concrete truth value that intuitively provides an upper bound on the truth value of the constraint's body, following the extension of classical constraints to the fuzzy case by Janssen et al.~\cite{JanssenSVC12}.
Formally, these rules and constraints take the forms
\begin{narrowgather}[5pt]
    \atom\lpif{}\body
    \qquad\text{and}\qquad
    c\lpif{}\body
\end{narrowgather}
\noindent with \mbox{$c\in\lset'$} and $\body$ a normal fuzzy formula.
The space of relevant interpretations is again the set $\Nint$ of all non-paraconsistent fuzzy interpretations \mbox{$\nint\colon\Atoms\to\lset$}.
Given a program $Q$ and a candidate interpretation $\nint$, the \define{program reduct} $Q_\nint$ is obtained by replacing every rule/constraint body $\body$ by its respective body reduct $\body_{\nint}$:
explicitly denoting the (non-negated/negated) atoms that occur in a body by $\body[p_1,\ldots,p_m,\flpnot p_{m+1},\ldots,\flpnot p_{n}]$, the \define{body reduct} $\body_{\nint}$ is then such that for all $x_1,\ldots,x_m\in\lset$, $\flpcontf{\body_{\nint}}(x_1,\ldots,x_m)\eqdef\flpcontf{\body}[x_1,\ldots,x_m,\flpnotf\nint(p_{m+1}),\ldots,\flpnotf\nint(p_n)]$.
As expected, then, a candidate \mbox{$\nint\in\Nint$} is a \define{stable model of $Q$} iff $\nint$ is the least model of $Q_{\nint}$ (with respect to the pointwise lifting of $\lleq$ to $\Nint$) and $\nint$ satisfies all constraints in $Q_{\nint}$ (where a constraint \mbox{$c\lpif{}\body$} is satisfied by $\nint$ iff \mbox{$\nfformeval{\nint}{\body}\lleq c$}).

Towards our reconstruction of that semantics, we first define how we translate programs in the syntax of Cornejo et al.~\cite{Cornejo2020extended} into our syntax of \Cref{def:syntax}.
\begin{definition}
    \label{def:cornejo-extended:translation}
    Let $Q$ be a fuzzy logic program over atoms $\Atoms$ according to~\cite{Cornejo2020extended}.
    For a rule $r=\atom\lpif{}\body\in Q$ for $\atom\in\Atoms$, define $\flp_r\eqdef\set{r}$;
    for a constraint $r=c\lpif{}\body\in Q$ for $c\in\lset'$, define its translation
    $\flp_r\eqdef\set{p_r\lpif{}\body,\; \flpneg p_r\lpif{}\flpnegf c }$, where $p_r$ is a fresh atom not occurring in $Q$.
    Finally, define $\flp_Q\eqdef\bigcup_{r\in Q}\flp_r$.
\end{definition}
Intuitively, the rule for $p_r$ records the actual truth value of the body $\body$;
the rule for $\flpneg p_r$ records the greatest possible truth value $c$ of $\body$ that is \emph{permitted} by $r$, by requiring a least truth value $\flpnegf c$ for $\flpneg p_r$.%
\footnote{We fix $\flpneg$ to be an involution such that $\flpnegf c\in V'$. This choice depends on the concrete $\lset'$; for the typical fuzzy logic case of $\lset'=[0,1]\cap\Q$, we can choose $\flpnegf x\eqdef 1-x$.}
The main idea to guarantee constraint satisfaction in the semantics is now to restrict models to the consistent ones and thus to filter out all interpretations where some constraint is violated.
Clearly, constraint violation is also the \emph{only} reason for a model to be filtered out, as there are no other rules with negated heads and thus no other potential sources for classical inconsistency.

Our main result of this section shows that our translation allows us to reconstruct their stable model semantics with our approximator from \Cref{def:approximator}.
Before proving the main result, we establish some useful prerequisites.
Similar to before, we lift $\nint\in\Nint$ to a paraconsistent interpretation $\fint_{\nint}$ by setting $\fint_{\nint}(\atom)\eqdef(\nint(\atom),\flpnegf\nint(\atom))$ for atoms occurring in $\flp_Q$ and $\fint_{\nint}(\atom)\eqdef(\nint(\atom),\least)$ otherwise.

\begin{lemma}
    \label{lem:cornejo-extended:body-evaluations}
    For any rule/constraint body $\body$ occurring in an extended fuzzy logic program $Q$ in the syntax of Cornejo et al.~\cite{Cornejo2020extended} and any non-paraconsistent fuzzy interpretation $\nint\in\Nint$, we have that for all $\fintalt\in\Fint$:
    \(
    \fnformeval{\fintalt}{\fint_{\nint}}{\body}=\fformeval{\fintalt}{\body_{\nint}}
    \).
    \begin{longproof}
        \begin{align*}
             & \pheq \fnformeval{\fintalt}{\fint_{\nint}}{\body}                                                                                                                                         \\
             & = \fnformeval{\fintalt}{\fint_{\nint}}{\body[\atom_1,\ldots,\atom_m,\flpnot\atom_{m+1},\ldots,\flpnot\atom_{n}]}
             & \text{(Notation)}                                                                                                                                                                         \\
             & = \flpcontf{\body}(\fformeval{\fintalt}{\atom_1},\ldots,\fformeval{\fintalt}{\atom_m},\flpnotf\fformeval{\fint_{\nint}}{\atom_{m+1}},\ldots,\flpnotf\fformeval{\fint_{\nint}}{\atom_{n}})
             & \text{(\Cref{def:approximator})}                                                                                                                                                          \\
             & = \flpcontf{\body}(\fformeval{\fintalt}{\atom_1},\ldots,\fformeval{\fintalt}{\atom_m},\flpnotf\nint(\atom_{m+1}),\ldots,\flpnotf\nint(\atom_{n}))
             & \text{(Definition of $\fint_{\nint}$)}                                                                                                                                                    \\
             & = \flpcontf{\body_{\nint}}(\fintalt(\atom_1),\ldots,\fintalt(\atom_m))
             & \text{(Definition of $\body_{\nint}$)}                                                                                                                                                    \\
             & = \fformeval{\fintalt}{\body_{\nint}[\atom_1,\ldots,\atom_m]}
             & \text{(Definition of $\fformeval{\fintalt}{\cdot}$)}                                                                                                                                      \\
             & = \fformeval{\fintalt}{\body_{\nint}}\qedhere
        \end{align*}
    \end{longproof}
\end{lemma}

\noindent Using this, we can again establish identity of operators.

\begin{lemma}
    \label{lem:cornejo-extended:operators-equal}
    For any extended fuzzy logic program $Q$ according to \cite{Cornejo2020extended} with translation $\flp_Q$, and
    non-paraconsistent fuzzy interpretation $\nint\in\Nint$ with counterpart $\fint_{\nint}$, we have \mbox{$\fst{\lpap[{\flp_Q}](\cdot,\fint_{\nint})}=\lpop[\flp_{Q_{\nint}}]$}.
    \begin{longproof}
        We have $Q_\nint=\set{\atom\lpif{}\body_{\nint}\guard \atom\lpif{}\body \in Q}\cup\set{c\lpif{}\body_{\nint}\guard c\lpif{}\body\in Q }$ by definition and therefore $\Atoms_Q\eqdef\set{p_r\guard r=c\lpif{}\body\in Q}=\Atoms_{Q_\nint}$.
        Furthermore, for all $\fintalt\in\Fint$ and $\atom\in\Atoms$,
        \begin{align*}
             & \pheq \fst{\lpap[{\flp_Q}]^+(\fintalt,\fint_\nint)}(\atom)                                                                                           \\
             & = \biglub \set{\fnformeval{\fintalt}{\fint_{\nint}}{\body} \guard p\lpif{}\body \in \flp_Q}               & \text{(\Cref{def:approximator})}         \\
             & = \biglub \set{\fnformeval{\fintalt}{\fint_{\nint}}{\body} \guard p\lpif{}\body \in Q}                    & \text{(Def.~$\flp_Q$)}                   \\
             & = \biglub \set{\fformeval{\fintalt}{\body_{\nint}} \guard \atom\lpif{}\body \in Q }                       & \text{(previous Claim)}                  \\
             & = \biglub \set{\fformeval{\fintalt}{\body_{\nint}} \guard \atom\lpif{}\body_{\nint} \in Q_{\nint}}        & \text{(Def.~$Q_{\nint}$)}                \\
             & = \biglub \set{\fformeval{\fintalt}{\body_{\nint}} \guard \atom\lpif{}\body_{\nint} \in \flp_{Q_{\nint}}} & \text{(Def.~$\flp_{Q_{\nint}}$)}         \\
             & = \lpop[\flp_{Q_{\nint}}]^+(\fintalt)(\atom)                                                              & \text{(\Cref{def:consequence-operator})}
        \end{align*}
        in the positive case and $\fst{\lpap[\flp_Q]^-(\fintalt,\fint_\nint)}(\atom)=\least$ in the negative case.
        For an atom $p_r$ arising from constraint $r=c\lpif{}\body^r$,
        \begin{align*}
             & \pheq \fst{\lpap[{\flp_Q}]^+(\fintalt,\fint_\nint)}(p_r)                                                                                               \\
             & = \biglub \set{\fnformeval{\fintalt}{\fint_{\nint}}{\body} \guard p_r\lpif{}\body \in \flp_Q}               & \text{(\Cref{def:approximator})}         \\
             & = \biglub \set{\fnformeval{\fintalt}{\fint_{\nint}}{\body^r} \guard c\lpif{}\body^r \in Q}                  & \text{(Def.~$\flp_Q$)}                   \\
             & = \biglub \set{\fformeval{\fintalt}{\body^r_{\nint}} \guard c\lpif{}\body^r \in Q }                         & \text{(previous Claim)}                  \\
             & = \biglub \set{\fformeval{\fintalt}{\body^r_{\nint}} \guard c\lpif{}\body^r_{\nint} \in Q_{\nint}}          & \text{(Def.~$Q_{\nint}$)}                \\
             & = \biglub \set{\fformeval{\fintalt}{\body^r_{\nint}} \guard p_r\lpif{}\body^r_{\nint} \in \flp_{Q_{\nint}}} & \text{(Def.~$\flp_{Q_{\nint}}$)}         \\
             & = \lpop[\flp_{Q_{\nint}}]^+(\fintalt)(p_r)                                                                  & \text{(\Cref{def:consequence-operator})}
        \end{align*}
        for the positive case, and for the negative case we obtain\/:
        \begin{align*}
             & \pheq \fst{\lpap[{\flp_Q}]^-(\fintalt,\fint_\nint)}(p_r)                                                                                           \\
             & = \biglub \set{\fnformeval{\fintalt}{\fint_{\nint}}{\body} \guard \flpneg p_r\lpif{}\body \in \flp_Q}           & \text{(\Cref{def:approximator})} \\
             & = \biglub \set{\fnformeval{\fintalt}{\fint_{\nint}}{\flpnegf c} \guard \flpneg p_r\lpif{}\flpnegf c \in \flp_Q} & \text{(Def.~$\flp_Q$)}           \\
             & = \flpnegf c                                                                                                    & \text{($p_r$ is unique)}         \\
             & = \biglub\set{ \fformeval{\fintalt}{\flpnegf c} \guard \flpneg p_r\lpif{}\flpnegf c \in \flp_{Q_{\nint}}}       & \text{($p_r$ is unique)}         \\
             & = \lpop[\flp_{Q_{\nint}}](p_r) \qquad \text{(\Cref{def:consequence-operator})}\qedhere
        \end{align*}
    \end{longproof}
\end{lemma}

\begin{theorem}
    \label{thm:cornejo-extended:correctness}
    Consider an extended fuzzy logic program $Q$ \cite{Cornejo2020extended} and its translation $\flp_Q$ obtained as above.
    An interpretation $\nint\in\Nint$ is a Cornejo et al.~\cite{Cornejo2020extended} stable model of $Q$ if and only if
    $(\fint_{\nint},\fint_{\nint})$ is a stable fixpoint of $\lpap[{\flp_Q}]$.
    \begin{longproof}
        The proof is similar to the proof of \Cref{thm:sakama-inoue:equivalence}:
        \begin{align*}
             & \phantom{\text{iff }} \nint \text{ is a stable model of } Q                                                                                                   \\
             & \text{iff } \nint \text{ is the least model of } Q_{\nint}                                             & \text{(Def.)}                                        \\
             & \text{iff } \fint_{\nint} \text{ is the least fixpoint of } \lpop[\flp_{Q_{\nint}}]                    & \text{(Def.)}                                        \\
             & \text{iff } \fint_{\nint} \text{ is the least fixpoint of } \fst{\lpap[{\flp_Q}](\cdot,\fint_{\nint})} & \text{(\Cref{lem:cornejo-extended:operators-equal})} \\
             & \text{iff } (\fint_{\nint},\fint_{\nint}) \text{ is a stable fixpoint of } \lpap[{\flp_Q}]             & \text{(Def.)}
        \end{align*}
        There is only one additional aspect, namely our translation of constraints $r=c\lpif{}\body\in Q$.
        It is clear that for any $\nint\in\Nint$, we have that $\nint$ satisfies $c\lpif{}\body$ iff $\fformeval{\fint_{\nint}}{\body}\lleq c$.
        On the other hand, assume that $(\fintalt,\fintalt)$ is a (stable) fixpoint of $\lpap[\flp_Q]$.
        Let $r=c\lpif{}\body^r\in Q$ be arbitrary.
        Since $(\fintalt,\fintalt)$ is a fixpoint of $\lpap[\flp_Q]$, we obtain that
        \[
            \fst{\fintalt(p_r)}
            =\fst{\lpap[\flp_Q](\fintalt,\fintalt)(p_r)}
            =\fst{\lpap[\flp_Q]^+(\fintalt,\fintalt)}(p_r)
            =\fformeval{\fintalt}{\body^r}
        \]
        and likewise
        \[
            \snd{\fintalt(p_r)}
            =\snd{\lpap[\flp_Q](\fintalt,\fintalt)(p_r)}
            =\fst{\lpap[\flp_Q]^-(\fintalt,\fintalt)}(p_r)
            =\flpnegf c
        \]
        whence we obtain the following\/:
        \begin{align*}
             & \phiff \fintalt(p_r) \text{ is consistent }                                                                 \\
             & \iff \fst{\fintalt(p_r)}\lleq\flpnegf\snd{\fintalt(p_r)}                                                    \\
             & \iff \fformeval{\fintalt}{\body^r}=\fst{\fintalt(p_r)}\lleq\flpnegf\snd{\fintalt(p_r)}=\flpnegf\flpnegf c=c \\
             & \iff \fintalt \text{ satisfies constraint } r\qedhere
        \end{align*}
    \end{longproof}
\end{theorem}

\section{New Contributions Based on Reconstructions}
\label{sec:new-contributions}

Beyond the previously reconstructed semantics, the approximator $\lpap$ yields an immediate semantic payoff.
First, we show that, with respect to  an \nflptext program $\flp$, the approximator $\lpap[\flp]$ naturally generalizes the approximator $\apKHS$ introduced by Kettmann et al.~\cite{KettmannHS25}.
Here we define the corresponding interpretation $\fint_{\nint}^+\in\Fint$ of a non-paraconsistent interpretation $\nint\in\Nint$ as $\fint_{\nint}^+ \eqdef (\nint,\set{\atom\mapsto\least\guard\atom\in\Atoms})$. %
\begin{proposition}
  \label{p:generalize-ijcai-ap}
  For any \nflptext $\flp$ and non-paraconsistent interpretations $\nint,\nintalt\in\Nint$ with paraconsistent counterparts $\fint_{\nint}^+,\fint_{\nintalt}^+\in\Fint$,
  we have $\fst{\lpap(\fint_{\nint}^+,\fint_{\nintalt}^+)}=\fint_{\fst{\apKHS(\nint,\nintalt)}}^+ $. %
  \begin{longproof}
    We first recall the definition of the approximator $\apKHS$.

    For a \nflptext program $\flp$, the associated symmetric approximator $\apKHS\colon\Nint^2\to\Nint^2$ introduced by Kettmann et al.~\cite{KettmannHS25} is defined via
    \[
      \fst{\apKHS(\nint,\nintalt)}(\atom)=
      \biglub \set{\flpweight\flpcontf{\flpand_i}\nfnformeval{\nint}{\nintalt}{\body} \guard \flpr \in \flp}
    \]
    where $(\flpand_i,\flpif{}_i)$ is an adjoint pair and $\nfnformeval{\nint}{\nintalt}{\body}$, the evaluation of $\body$ under interpretation pair $(\nint,\nintalt)\in\bil[\nint]$, is defined via structural induction with notable base cases
    $\nfnformeval{\nint}{\nintalt}{\atom}\eqdef\nint(\atom)$ and
    $\nfnformeval{\nint}{\nintalt}{\flpnot\atom}\eqdef\flpnotf\nintalt(\atom)$,
    and straightforward inductive case $\nfnformeval{\nint}{\nintalt}{f(\varphi_1,\ldots,\varphi_n)}\eqdef\flpcontf{f}\!\left(\nfnformeval{\nint}{\nintalt}{\varphi_1},\ldots,\nfnformeval{\nint}{\nintalt}{\varphi_n}\right)$.

    \begin{claim}
      \label{c:body-eval-equal}
      For any normal fuzzy body formula $\body$, we have
      \[
        \nfnformeval{\nint}{\nintalt}{\body}=\fnformeval{\fint_\nint^+}{\fint_\nintalt^+}{\body}
      \]
      \begin{claimproof}
        \begin{align*}
           & \pheq \nfnformeval{\nint}{\nintalt}{\body}                                                                                                              \\
           & =  \nfnformeval{\nint}{\nintalt}{\body[p_1,\ldots,p_m,\flpnot p_{m+1},\ldots,\flpnot p_{n}]}                                                            \\
           & = \flpcontf{\body}(\nint(p_1),\ldots,\nint(p_m),\flpnotf\nintalt(p_{m+1}),\ldots,\flpnotf\nintalt(p_n))                                                 \\
           & = \flpcontf{\body}(\fint_{\nint}^+(p_1)_1,\ldots,\fint_{\nint}^+(p_m)_1,\flpnotf\fint_{\nintalt}^+(p_{m+1})_1,\ldots,\flpnotf\fint_{\nintalt}^+(p_n)_1) \\
           & = \fnformeval{\fint_\nint^+}{\fint_\nintalt^+}{\body} \tag*{\(\triangle\)}
        \end{align*}
      \end{claimproof}
    \end{claim}

    For simplicity, we consider only weight-free rules.
    \begin{description}
      \item
            Then we have the following identities:
            \begin{align*}
               & \pheq \fst{\lpap(\fint_{\nint}^+,\fint_{\nintalt}^+)}^+(\atom)                                                 \\
               & = \biglub \set{\fnformeval{\lint}{\uint}{\body} \guard \eflprp \in \flp}   & \text{(Def.)}                     \\
               & = \biglub \set{\nfnformeval{\nint}{\nintalt}{\body} \mid \eflprp \in \flp} & \text{(\Cref{c:body-eval-equal})} \\
               & = \fst{\apKHS(\nint,\nintalt)}(\atom)                                      & \text{(Def.)}
            \end{align*}
      \item
            Since $\flp$ is a $\flpneg$~free program, $\fst{\lpap(\fint_{\nint}^+,\fint_{\nintalt}^+)}^-(\atom) = \least$ for any $\atom\in\Atoms$ and interpretation $\fint_{\nint}^+,\fint_{\nintalt}^-\in\Fint$, hence $\fst{\lpap(\fint_{\nint}^+,\fint_{\nintalt}^+)}=(\fst{\apKHS(\nint,\nintalt)},\set{\atom\mapsto\least\guard\atom\in\Atoms})$.\qed
    \end{description}
  \end{longproof}
\end{proposition}

In particular, this result implies that the well-founded semantics defined by Loyer and Straccia~\cite{LoyerS09} arises as a special case of our fuzzy well-founded semantics.
More generally, our framework provides a paraconsistent well-founded semantics for \eflptext, extending the semantics of Loyer and Straccia~\cite{LoyerS09} to paraconsistency, and lifting the paraconsistent semantics of Sakama~\citeyearpar{sakama1992extended} from the crisp to the fuzzy setting.
It also induces paraconsistent stable models for \eflptext, extending the work of
Sakama and Inoue~\cite{SakamaI95},
Cornejo et al.~\cite{Cornejo2020extended}, and Janssen et al.~\cite{JanssenSVC12}.

\section{Discussion}

We already discussed relations with a host of related work in the reconstructive sections (Sections \ref{sec:reconstructions-crisp} and \ref{sec:reconstructions-fuzzy}).
Here, we survey other related work.
One line of work on extended logic programs that we did not reconstruct is that done by Alferes, Damásio, Pereira, and Alc{\^a}ntara~\cite{alferes1995logic,damasio1995model,alcantara2005encompassing}.
They introduce the \emph{well-founded semantics for extended logic programs}.
Their semantics is a straightforward adaption of the well-founded semantics for normal logic programming \cite{vanGelder88}, but is based on what they call the \emph{principle of coherence}:
``believing $\flpneg p$ entails believing $\flpnot p$''.
By double negation, this also means that believing $p$ entails believing $\flpnot \flpneg p$.
In the presence of contradictions, the AFT-perspective we take (and, as shown by the reconstructions, also underlies other approaches) is incompatible with the principle of coherence.
Indeed, as we only consider pairs $(\lint,\uint)$ s.t.\ $\lint\tleq \uint$, any time we believe $p$ and $\flpneg p$ (i.e.\ $p,\flpneg p\in \lint$), we also believe in the non-falsity of $p$ and $\flpneg p$ (i.e.\ $p,\flpneg p \in \uint$), which precludes us from believing $\flpnot \flpneg p$ respectively $\flpnot p$.
On the other hand, Alc{\^a}ntara et al.~\cite{alcantara2005encompassing} need some additional, restrictive assumptions on their negation functions to make full use of the coherence principle:
both classical and default negations must be involutions (self-inverse bijections), and furthermore they must commute.
In our framework, we need no such assumptions and can thus also accommodate Gödel negation (which is not involutive).

Another related work is that of Killen and You~\cite{killenEliminatingUnintendedStable2023}, who use tetralattices to propose an alternative definition of stable revision.
Essentially, they use tetralattices to obtain pairs $(T,F,U,P)$, where $T$ gives necessary truth, $F$ well-established falsity, $U$ possible falsity and $P$ possible truth.
On the other hand, we consider pairs of pairs $((\lint_1,\lint_2),(\uint_1,\uint_2))$ where $\uint_1$ and $\uint_2$ [$\lint_1$ and $\lint_2$] represent the possible [necessary] truth of $p$ respectively $\flpneg p$.

We also provide a proof-of-concept implementation in ASP, using the \emph{meta-programming} capabilities of clingo~\cite{kaminski2023build}. We restricted our implementation to fixpoint and Kripke-Kleene semantics for crisp \eflpstext.
\iflong
    In meta-programming we encode a grounded logic program as a set of facts called a \define{reification}.
    The semantics is encoded in a logic program called a \define{meta-program}, and applied by feeding both the reification and meta-program into clingo for grounding and solving.
    Since clingo's grounder gringo performs some simplifications and defines away strong negation to a constraint, we had to introduce a predicate \verb|snot| which encodes strong negation (instead of using clingo's strong negation \verb|-|).
\fi
More details can be found in the online repository at \url{https://github.com/JeroenSpaans/sningo}.

Janssen et al.~\cite{JanssenSVC12} showed that several common extensions of normal fuzzy logic programs, such as constraints and extended aggregators, can be simulated by normal rules while preserving the fuzzy answer set semantics.
For the extension allowing strong negation, however, the semantics is based on a slightly different notion of modelhood, using the approach of Van Nieuwenborgh et al.~\cite{VanNieuwenborgh2007b}.
In this setting, two special aggregators are employed:
one to determine the degree to which an interpretation $\fint$ is consistent (by aggregating the consistency values induced by an associated consistency function across all atoms), and another to determine the degree to which $\fint$ is a model of $\flp$ (by aggregating the satisfaction degrees of all rules of $\flp$).
We aim to generalize this approach in future work.
Other planned future work includes the generalization of this work to extended or fuzzy logic programs with disjunction, using non-deterministic AFT~\citep{HeyninckAB24}.
Furthermore, we see several potential applications of extended fuzzy logic programs, e.g.\ in law, where conflicting obligations have to be combined with fuzzy concepts~\citep{atkinson2023angelic}, or in neuro-symbolic AI, where fuzzy logic is often used to integrate logical rules in a loss function~\citep{sen2022neuro}, and where hallucinations might cause inconsistencies.

\subsubsection{\ackname}

This work is partly supported by BMFTR (Federal Ministry of Research, Technology, and Space) in DAAD project 57616814 (\href{https://secai.org/}{SECAI}).%
\subsubsection{\discintname}
The authors have no competing interests to declare that are relevant to the content of this article.

\bibliographystyle{splncs04}
\bibliography{references}

\end{document}